\documentclass[11pt, letterpaper]{article}

\usepackage{graphicx}
\usepackage{amssymb}
\usepackage{float}
\usepackage{setspace}
\usepackage{epstopdf}
\usepackage{color}
\usepackage[letterpaper,left=1in,right=1in,top=1in,bottom=1in]{geometry}
\usepackage{multirow}
\usepackage{longtable}
\usepackage{rotating}

\usepackage{makeidx}
\usepackage{ragged2e}
\usepackage{caption}
\usepackage{hyperref}
\usepackage{mwe}
\usepackage{mathrsfs}
\usepackage{verbatim}
\usepackage{subfigure}
\usepackage{lineno}
\usepackage{algorithm,algcompatible,amsmath}

\algnewcommand\INPUT{\item[\textbf{Input:}]}%
\algnewcommand\OUTPUT{\item[\textbf{Output:}]}%

\newcommand{\m}{\mathbb}

\title{\Large\bf Adaptive Model Reduction and State Estimation of Agro-hydrological Systems}
\author{
\centerline{\normalsize Soumya Ranjan Sahoo$^{a}$, Jinfeng Liu$^{a,}$\thanks{Corresponding author: J. Liu. Tel: +1-780-492-1317. Fax: +1-780-492-2881. Email: jinfeng@ualberta.ca.}}
\vspace{5mm}\\
\centerline{\small $^{a}$Department of Chemical \& Materials Engineering, University of Alberta,} \\
\centerline{\small Edmonton, AB T6G 1H9, Canada}}

\allowdisplaybreaks

\begin{document}

\date{}
\maketitle
\doublespacing

\begin{abstract}
Closed-loop irrigation can deliver a promising solution for precision irrigation. The accurate soil moisture (state) estimation is critical in implementing the closed-loop irrigation of agro-hydrological systems. In general, the agricultural fields are high dimensional systems. Due to the high dimensionality for a large field, it is very challenging to solve an optimization-based advanced state estimator like moving horizon estimation (MHE). This work addresses the aforementioned challenge and proposes a systematic approach for state estimation of large agricultural fields. We use a non-linear state-space model based on discretization of the cylindrical coordinate version of Richards equation to describe the agro-hydrological systems equipped with a central pivot irrigation system. We propose a structure-preserving adaptive model reduction method using trajectory-based unsupervised machine learning techniques. Furthermore, the adaptive MHE algorithm is developed based on an adaptive reduced model. The proposed algorithms are applied to a small simulated field to compare the performance of adaptive MHE over original MHE. Finally, the proposed approach is applied to a large-scale real agricultural field to test the effectiveness and superiority to address the current challenges. Extensive simulations are carried out to show the efficiency of the proposed approach. 
\end{abstract}


\section{Introduction}
The scarcity of freshwater is one of the most significant global risks \cite{global_2015}. Agricultural irrigation consumes about 70\% of the total freshwater usages \cite{wastewater_2017}. However, the average irrigation water usage efficiency of the current irrigation methods is around 50\% to 60\% \cite{lozoya_model_2014}. To feed the growing population and to manage the water crisis, the important step is to increase the water-use efficiency. 

One of the reasons for the poor water-use efficiency in the current irrigation practice is the decision-making process. In general, the irrigation prescription is made in an open-loop manner; that is, the irrigation amount is determined based on the farmer's experience instead of the actual field condition (e.g., soil moisture). It is well recognized that open-loop control is not precise. The use of open-loop irrigation often leads to excessive or insufficient irrigation. One promising solution is to close the decision-making loop and form a closed-loop irrigation system. It is expected that the water use efficiency can be increased significantly by implementing a closed-loop irrigation system, where the controller makes decisions based on real-time field information. One of the vital aspects of closed-loop irrigation is to know the field condition to make a decision. The field condition can be measured based on soil moisture sensors, but the soil moisture sensors cannot be placed everywhere in the field. So for a large agricultural field, one of the feasible solutions is to estimate the field condition or system states using a limited number of sensors. 

The data assimilation or estimation of the soil moisture has been extensively studied in the past using popular methods like extended Kalman filter (EKF) \cite{reichle_extended_2002, lu_dual_2011}, 
ensemble Kalman filter (EnKF) \cite{lannoy_state_2007, wu_feasibility_2011, pasetto_ensemble_2012, erdal_importance_2015, kurtz_terrsysmppdaf_2016, zhang_state_2017}, particle filter \cite{montzka_hydraulic_2011}, and moving horizon estimation (MHE) \cite{bo_parameter_2020,bo_simultaneous_2020, bo_decentralized_2020}. The major challenge in estimation using PF, EKF and EnKF methods is that it cannot handle the constraints explicitly. In general, we do not have enough information about actual initial system states, so while designing the EKF and EnKF, we may need to provide a high initial covariance matrix. The high covariance may take some states to the extreme pressure head condition (dry or saturated region) and solve the agro-hydrological system using Richards equation at extreme pressure head conditions faces numerical challenges \cite{farthing_numerical_2017}. In Erdal et al. \cite{erdal_importance_2015}, it has also been shown that under extreme pressure head conditions, the EnKF becomes unstable. 
Another limitation of EKF and EnKF methods is the assumption of Gaussian noise. In the presence of non-Gaussian noise, the estimation performance may deteriorate. MHE can handle the above-discussed problems. In our previous works, we have attempted to use the MHE for state and parameter estimation for both 1D and 3D system \cite{bo_parameter_2020, bo_simultaneous_2020, bo_decentralized_2020}. But there are still some limitations, such as it applies to a 1D system or a small 3D system, which can be approximated to a 1D system. The challenges arise when solving an optimization problem for a large agricultural field with heterogeneous soil type distribution with non-uniform irrigation amount. 

An agro-hydrological system can be modeled using the Richards equation, a three-dimensional PDE equation. Richards equation assumes the local equilibrium present in the soil water content and soil water potential. The finer resolution grid (few centimetres to few meters) is often required to solve the Richards equation not only to ensure numerical stability but also to satisfy the local equilibrium assumption \cite{babaeian_ground_2019}. The finer resolution makes the number of states very high in the range of ($10^4-10^8$). The increased system order makes the estimation and controller design very challenging, and if MHE is used, it makes the optimization problem intractable. Model reduction is one of the widely accepted techniques to handle such types of high-dimensional systems. A few popular model reduction techniques are proper orthogonal decomposition (POD), optimal Hankel norm reduction, balanced truncation methods \cite{antoulas_approximation_2005}. However, these methods do not preserve the network topology or physical meaning of the states. Thus these methods may not be useful while applying state constraints in the design of the MHE. 
In \cite{ishizaki_model_2014,cheng_graph_2016,  cheng_gramian-basedmodel_2019,sahoo_optimal_2019} structure-preserving cluster based model reduction is proposed. In these methods, the obtained clusters do not overlap, and the states in each cluster are aggregated to one single reduced-order state. In each cluster, the dynamic properties are almost the same. So the reduced-order model preserves the essential dynamic properties as well as the physical topology. However, these methods are applied to linear systems, and one reduced model is constructed offline for the whole simulation time. 

In general, the reduced model captures the dynamics of the actual system by a low-dimensional manifold. In some cases, the solution of the actual system is scattered all over the high dimensional space, or it changes over time. In such cases, one reduced model may not represent the high dimensional manifold of the actual system or requires a very high dimensional reduced space. 
In such scenarios, the adaptive model reduction is one of the promising solution. In \cite{peherstorfer_dynamic_2015, peherstorfer_online_2015}, the POD based adaptive model reduction is proposed. In our previous work \cite{sahoo_dynamic_2020}, we have proposed adaptive model reduction for sensor placement using the reduced model at different linearized points. 
 
In this work, we propose a novel structure-preserving adaptive model reduction method for state estimation of large-scale agro-hydrological systems. The main idea is to construct different reduced models at different periods and use the reduced models to estimate the states of the original model. First, the system trajectories are generated using the original non-linear model. Then the unsupervised machine learning clustering approach is used to find the states having similar trajectories and put them in one cluster. Next, the projection-based reduced model is constructed using the clustering information. As the model may change over time, the changing boundary condition is proposed to move from one model to another. 
Similar to the idea of adaptive model reduction, the adaptive moving horizon estimation is proposed. The challenges arise because the model changes over time. The proposed adaptive reduced-order MHE considers the information exchange of reduced models and estimates the original system's states. 
The effectiveness of the proposed adaptive MHE is compared with the original MHE for a small system. We also apply the proposed adaptive MHE to the real agricultural field located in Lethbridge, Alberta, with the heterogeneous soil type and with the central pivot irrigation system. The main advantages of using the adaptive reduced-order system are as follows: 
\begin{enumerate}
    \item The adaptive model reduction approach considers the local dynamics of the system, so the number of states of the adaptive reduced model is lesser than the global model reduction models. 
    \item In MHE, the number of decision variables is the number of states times the horizon length. For a  large agricultural field, the number of states ($x$) is very high so as the decision variables and make the optimization problem intractable. 
    In the proposed adaptive reduced model, the number of reduced states ($\xi$) is way lesser than the actual system states ($\xi<<x$). Thus the number of decision variables in the adaptive MHE is lesser and makes the optimization problem tractable. 
    \item The degree of observability of the adaptive MHE is higher than the original MHE, which signifies that using the same number of measurements, the adaptive MHE can estimates states faster than the original MHE. \item The adaptive model reduction is also helpful while estimating using EKF or EnKF. For EKF, instead of calculating the huge jacobian matrix, the reduced-order jacobian matrix can be used. Similarly, the large matrix calculations can be avoided while using EnkF. 
    \item The adaptive reduced-order model is also helpful in other control problems like sensor placement and controller design. 
\end{enumerate}

The remainder of this work is organized as follows. Section 2 discusses the model and numerical schemes of the agro-hydrological system. The algorithm for adaptive model reduction is presented in section 3. Section 4 describes the adaptive moving horizon estimation. Section 5 presents the results of the proposed adaptive MHE to a small field. In section 6, the proposed adaptive model reduction and estimation are applied to an actual agricultural field under different scenarios.

\section{System description and problem formulation}
\subsection{System description}

\begin{figure}
\centering
\includegraphics[width=0.7\textwidth]{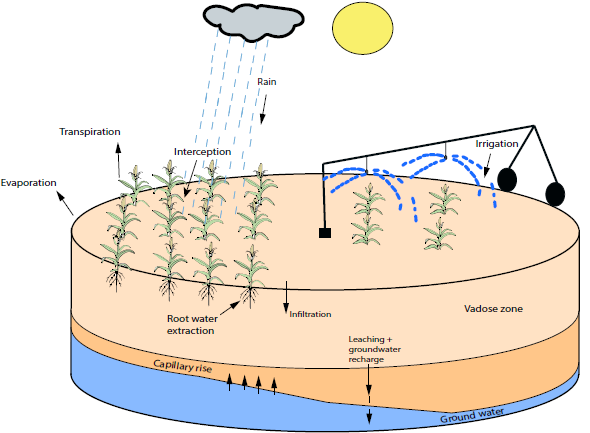}
\caption{Schematic of an agro-hydrological system \cite{agyeman_soil_2020}}
\label{fig:agro}
\end{figure}

An agro-hydrological system characterizes the hydrological cycle among soil, water, atmosphere and crop. A schematic of the agro-hydrological system considered is shown in Figure~\ref{fig:agro}. The water inflows to the agro-hydrological system are irrigation, precipitation and the system outflows are evaporation, transpiration and ground water discharge. 
The dynamics of the water flows in the agro-hydrological system are governed by the Richards equation as follows \cite{richards_capillary_1931}:

\begin{equation}
{\dfrac{\partial \theta}{\partial t}=c(h)\dfrac{\partial h}{\partial t}=\nabla \cdot(K(h)\nabla(h+z))+S(h,z)}
\label{eq:richards}
\end{equation}
where $h$ [m] is the field water pressure head, $\theta$ [m$^{3}$m$^{-3}$]  is the field water soil moisture content, $c(h)$ [m$^{-1}$] is the soil water capacity, $K(h)$ [ms$^{-1}$] is the hydraulic conductivity, $z$ [m] is the vertical coordinate, $S(h,z)$ [m$^{3}$m$^{-3}$s$^{-1}$] is the sink term. 

The relation between the hydraulic conductivity $K(h)$ and pressure head $h$ is characterized as follows \cite{mualem_new_1976}:

 \begin{equation}
K(h)=
\begin{cases}
K_{sat}S_e^\lambda[1-(1-(S_e^{\frac{1}{m}})^m]^2, & S_e < 1\\
K_{sat}, & S_e \geq 1
\end{cases}
\end{equation}
\vspace{-2mm}

\noindent where $S_e = 1+(-\alpha h)^n$, $\lambda, \alpha, m, n$ are the shape factors and $K_{sat}$ [ms$^{-1}$] is the saturated hydraulic conductivity. The capillary capacity $c(h)$ is computed as follows \cite{van_genuchten_closed-form_1980}:
\begin{equation}
c(h) =
\begin{cases}
(\theta_{s}-\theta_{r})\alpha n(1-(\frac{1}{n}))(-\alpha h)^{n-1}(1+(-\alpha h)^n)^{-(2-(1/n))}, & h < 0 \\
S_{r}, & h \geq 0
\end{cases}
\end{equation}
where $\theta_s$ [m$^{3}$m$^{-3}$] denotes the saturated soil water content, $\theta_r$ [m$^{3}$m$^{-3}$] denotes the residual water content and $S_{r}$ [m$^{-1}$] the specific storage coefficient of the porous medium under positive pressure.
\noindent
The relation between moisture content ($\theta$) and the pressure head ($h$) can be expressed as \cite{van_genuchten_closed-form_1980}:
\begin{equation}
\theta=
\begin{cases}
(\theta_{s}-\theta_{r})(1+(-\alpha h)^n)^{-(1-(1/n))}+\theta_{r}, & h< 0 \\
\theta_{s}, & h \geq 0
\end{cases}
\end{equation}
The sink term $S(h,z)$ in (\ref{eq:richards}) is the root water extraction rate. In this work, the optimum water uptake is assumed and it is calculated as follows \cite{allen_crop_nodate}: 
\begin{equation}
    S_{\text{max}}(h,z) = \frac{ETP_p}{L}
\end{equation}
where $ETP_p$ [ms$^{-1}$] is the potential evaporation rate and $L$ [m] is the rooting depth. The potential evaporation rate $ETP_p$ is computed by:  
\begin{equation}
    ETP_p = K_cPET
\label{eq:ET}
\end{equation}
where $PET$ [ms$^{-1}]$ is the reference evaporation rate which can be calculated based on the Penmon-Moneith equation \cite{allen_crop_nodate} and $K_c$ [-] is the crop coefficient. 

In this work, we consider that the agricultural field is equipped with a center pivot irrigation system. A center pivot irrigation system rotates across the field around a fixed pivot at the center of the field and irrigates in a circular manner. In order to account for the circular movement of the center pivot irrigation system, the Richards equation in (\ref{eq:richards}) is expressed in the cylindrical coordinates as follows \cite{agyeman_soil_2020}: 

\begin{equation}
    c(h)\dfrac{\partial h}{\partial t}= \frac{1}{r}\frac{\partial}{\partial r}\Bigg[rK(h)\frac{\partial h}{\partial r}\Bigg]+
    \frac{1}{r}\frac{\partial}{\partial \theta}\Bigg[\frac{K(h)}{r}\frac{\partial h}{\partial \theta}\Bigg]+
    \frac{\partial}{\partial z}\Bigg[K(h)\Bigg(\frac{\partial h}{\partial z}+1\Bigg)\Bigg]+S(h,z)
\label{eq:cyl_richards}
\end{equation}

The three-dimensional agro-hydrological model in (\ref{eq:cyl_richards}) is a nonlinear partial differential equation, which renders the problem difficult to solve analytically. In this work, we apply the explicit finite difference method to discretize the Richards equation in (\ref{eq:cyl_richards}). Note that spatial discretization of the model is performed, such that a continuous-time state-space model is established as in the following form:
\begin{equation}
\begin{aligned}
\dot{x}(t)&= f(x(t), u(t))
\end{aligned}
\label{eq:nonlinear}
\end{equation}
where $x(t)\in \mathbb{R}^{N_x}$ denotes the states vector  representing the pressure head value at each discretized node of total size $N_x$ and $u \in \mathbb{R}^{N_u}$ represents the input vector containing $N_u$ irrigation values applied at each surface discretized node. As the input (irrigation amount) is applied to each surface node, it is incorporated in the system surface boundary condition. The surface boundary condition is characterized by Neumann boundary condition as follows:
\[
\frac{\partial{h}}{\partial{z}}\Bigg|_{r,\theta,z= Z_s} = -1 - \frac{u(t)}{K(h)}
\]
where $h$ is the pressure head, $u(t)$ is the input to the system, $K(h)$ is the hydraulic conductivity and $Z_s$ is the length of the soil column. The bottom boundary condition is specified as free drainage.

\subsection{Problem formulation}
As described in the previous subsection, the dimension of $u$ is the same as the number of surface nodes. However, at a specific time, only the elements in $u$ that align with the central pivot may not be zero, and the other elements are zero. This imposes a time-varying constraint on $u$ as follows:
\begin{equation}
    \mathscr{U}_{lb}(t) \leq u(t) \leq \mathscr{U}_{ub}(t)
\end{equation}

We consider tensiometer sensors to measure the soil pressure head values (states) at a few selected locations on the field in this work. The number of sensors ($N_y$) are significantly smaller than the number of states ($N_x$) of the system ($N_y << N_x$). At every sampling time, the tensiometer sensors provide the soil pressure head value for the corresponding sensor locations. 

Therefore, a continuous time state-space model with measurements is considered as follows: 
\begin{subequations}
\begin{align}
\dot{x}(t)&= f(x(t), u(t)) + w(t) \label{eq:ss_w} \\
y(t)&  = Cx(t) +v(t)
\end{align}
\label{eq:statespace}%
\end{subequations}
where $y(t)\in \mathbb{R}^{N_y}$ denotes the soil pressure head measurements at the sensor nodes, $w(t)\in \mathbb{R}^{N_x}$ is the additive system noise and $v(t)\in \mathbb{R}^{N_y}$ denotes the measurement noise. 

The objective is to estimate the soil pressure head at each node in the field using sensor measurements for a three-dimensional field with a central pivot irrigation system and heterogeneous soil type.  

\section{Proposed adaptive model reduction} \label{section3}
Let us consider the three-dimensional Richards equation in (\ref{eq:cyl_richards}). As discussed in the introduction part that the finer resolution discretization is required to solve the Richards equation. Thus the number of states in (\ref{eq:statespace}) becomes very high, and it is challenging to perform the optimization-based state estimation. Model reduction is one of the solutions to handle the problem. In this section, we propose a structure-preserving adaptive model reduction method to address this issue. 

There are mainly two challenges associated with the classical model reduction algorithms: 1) The reduced model does not preserve the physical significance of the states. 
Given that the goal of the model reduction is to perform state estimation based on MHE, it is preferred that the reduced-model preserves the system network topology while applying state constraints to the MHE. 
2) In general, the reduced model tries to capture the actual system trajectory using low-dimensional space. If the trajectories of the actual system are distributed sparsely in higher dimension space, it is preferable to change the reduced model at different time periods. 

These two challenges motivate us to construct a structure-preserving adaptive model reduction method based on the unsupervised machine learning clustering technique. The motivation to use the adaptive model reduction in the agro-hydrological system is illustrated using a simple example in below section \ref{sec:moti}. 
\subsection{Motivating example}
\label{sec:moti}
\begin{figure}
\centering
\includegraphics[width=1\textwidth]{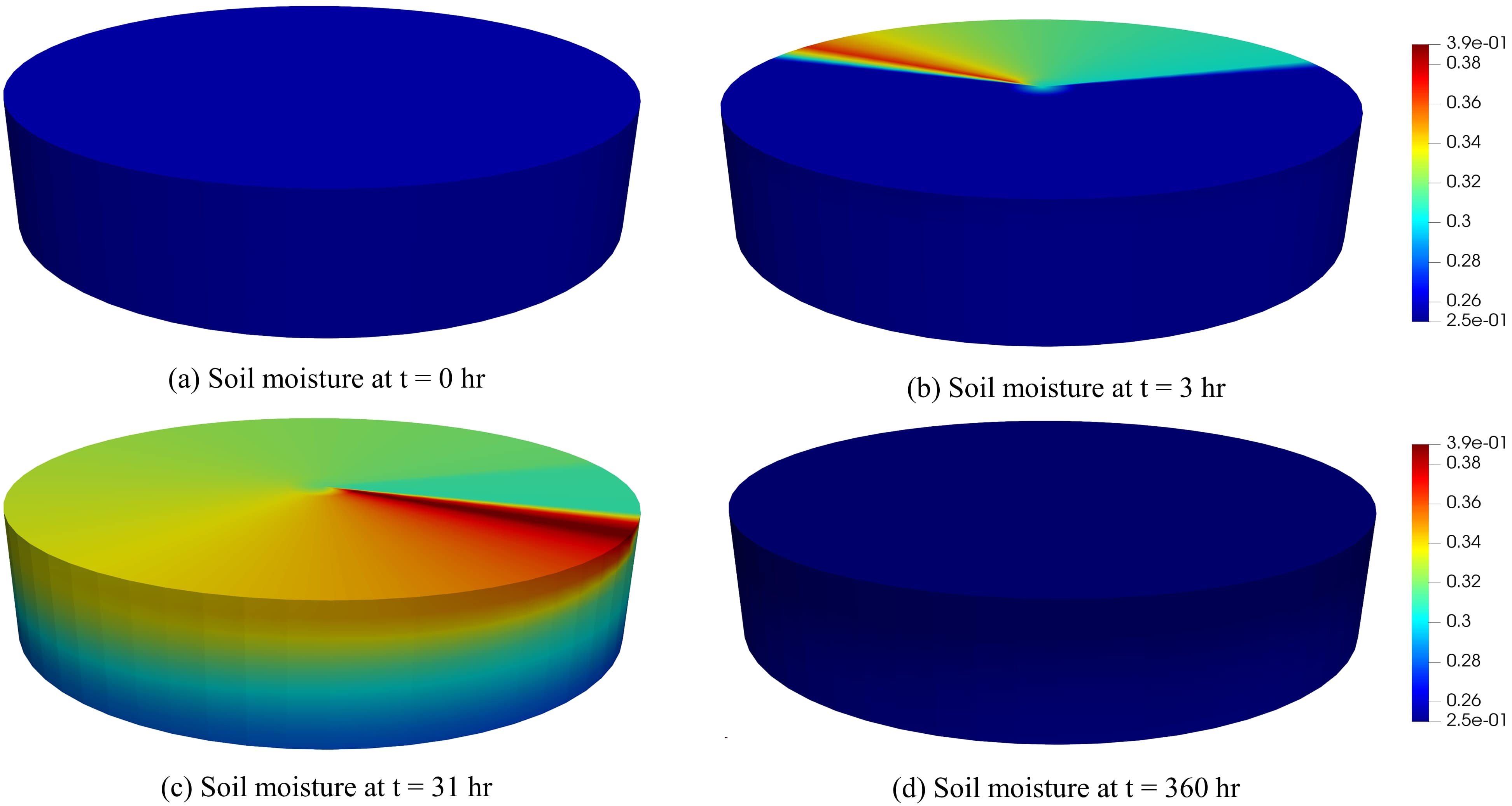}
\caption{Motivational example for adaptive model reduction}
\label{fig:motivation}
\end{figure}

 In the agro-hydrological system, the input usually follows a non-periodic way. For instance, the central pivot as an irrigation system circularly puts water (Figure \ref{fig:agro}). So the system dynamics may change substantially both in spatial and temporal directions from one node to another. For example, consider a field with uniform initial conditions and homogeneous soil types. The central pivot takes 8 hrs to complete one circle. The total simulation time period is 15 days. For the first two days, there is continuous irrigation, and the irrigation amount is zero for the other remaining days. From Figure \ref{fig:motivation}(a) at $t = 0 $ hr, we can observe that the soil moisture all over the field is constant, and the system can be represented by only one node. When the irrigation starts, the system dynamics change spatially and temporally, so more nodes are required to capture the system dynamics. From Figure \ref{fig:motivation}(b) and \ref{fig:motivation}(c), we notice that at the $3^{rd}$ hr, the system dynamics is lesser as compared to the $31^{th}$ hr. After we stop the irrigation, some of the state values gradually become the same, and the system dynamics reduce again (Figure \ref{fig:motivation}). Thus few number nodes are required to capture the dynamics. So if we represent the system with only one global reduced model, we may need to use a higher-order reduced model to capture all the system dynamics. Hence it is preferable to change the reduced model over time based on the system's trajectory. In other words, we can interpret the adaptive reduced model as an adaptive meshing technique in which the system mesh becomes finer where the system dynamics are high. 
 
 \subsection{Proposed method}
\begin{figure}
\centerline{\includegraphics[width=1\columnwidth]{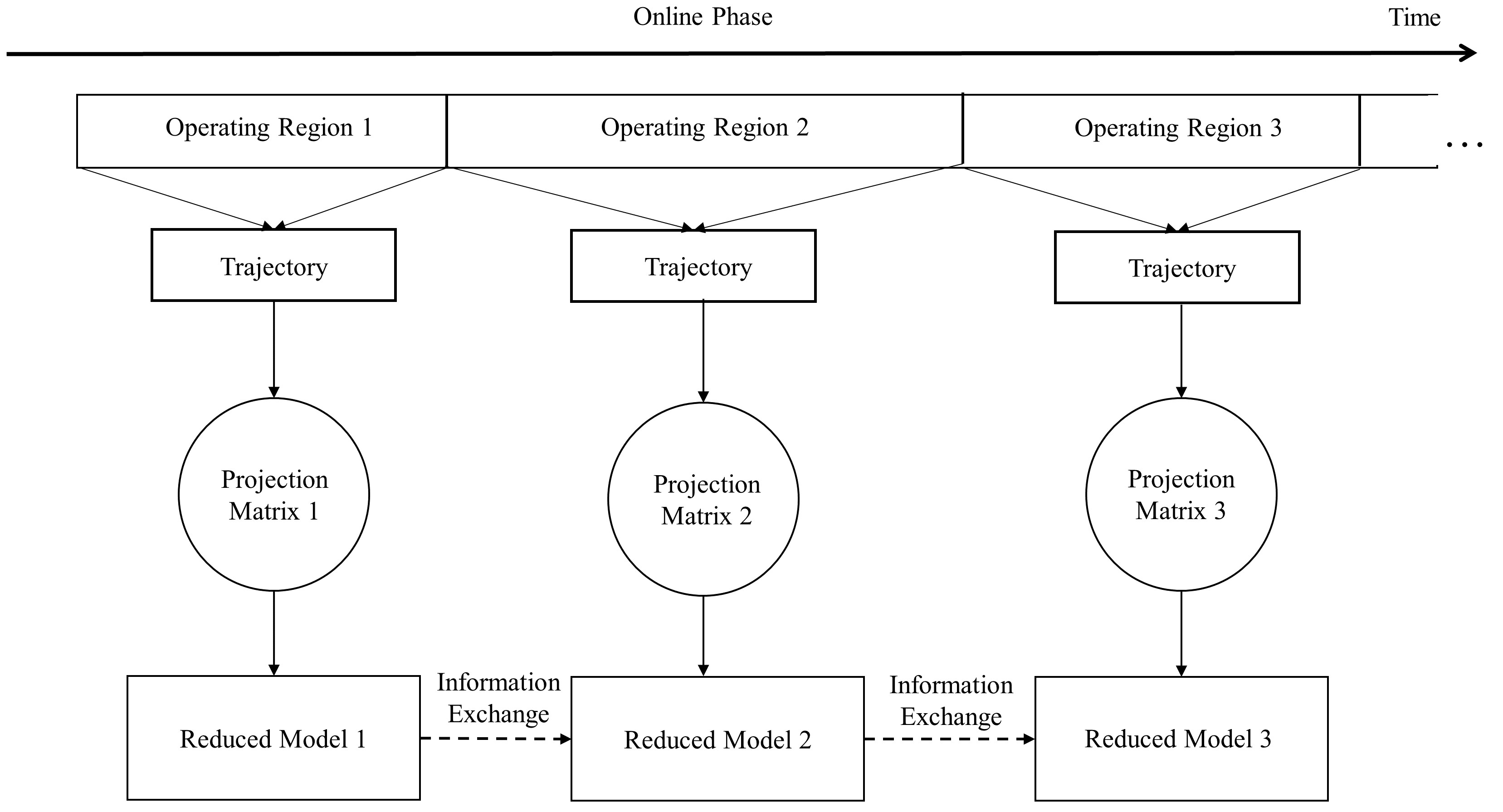}}
\caption{Steps to calculate adaptive reduced model.} \label{fig:steps}
\end{figure}

\begin{figure}
\centerline{\includegraphics[width=1\columnwidth]{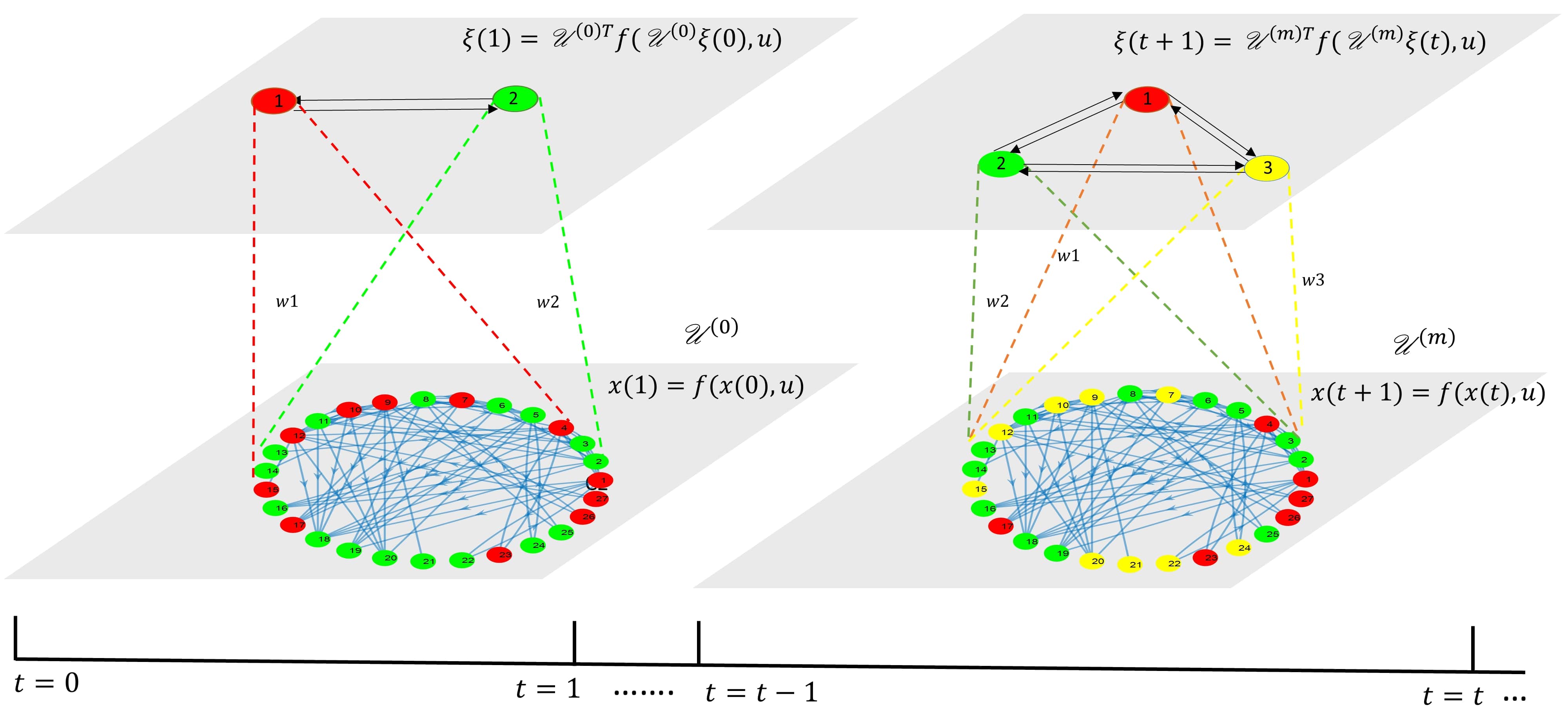}}
\caption{Illustration of adaptive model reduction.} 
\label{fig:dmr}
\end{figure}

 The key steps to calculate the adaptive model reduction are shown in Figure \ref{fig:steps}. The figure shows that the first step is to divide the total time into different operating regions. In the second step, the trajectories are generated for each operating region. In the third step, the clusters and the projection matrix are generated based on the similarity between the trajectories. In the same step, using the projection matrix, the reduced-order model is obtained. In the final stage, the information exchange between the reduced-order model is considered. 
 The illustration of the proposed adaptive model is shown in Figure \ref{fig:dmr}. The details of each step in the proposed algorithm are described as follows. 
 
\subsubsection{Step 1: operating region calculation} 
The idea in this step is to divide the total time period of the system into different operating regions based on input sequence or system trajectories.
Let us consider the motivating example again. While there is irrigation for the first two days, the system dynamics change rapidly every sampling time. So in the first two days, we may divide the time with finer size operating regions. After the irrigation stops, the system dynamics decreases gradually, so the coarse size operating regions may be used. For each operating region, one reduced model is constructed. Let us consider $\mathscr{R} = \{\mathscr{R}_1,\mathscr{R}_2, \dots, \mathscr{R}_s\}$ as the collection of operating regions, which has the following property: $\mathscr{R}_1 \cup \mathscr{R}_2 \cup\ldots \cup \mathscr{R}_s =\mathscr{T}$, where $\mathscr{T}= [0, T_f]$, $T_f$ is total operating time, $s$ is the total number of operating regions and $\mathscr{R}_1 \cap \mathscr{R}_2 \cap\ldots \cap \mathscr{R}_s \neq \emptyset$ may hold. Things to note that the cardinality of each set of operating regions may be different.

\subsubsection{Step 2: trajectory generation} In this step, the state snapshots are generated for each operating region. Let us consider the nonlinear system in (\ref{eq:nonlinear}). Based on a prescribed input, simulate the original nonlinear system and capture the trajectories in each operating region. For each operating region ($\mathscr{R}_m \in \mathscr{R}$),  the trajectories can be generated as follows: 

    \[\mathscr{X}_m =[x(t_{0,m}) ~x(t_{1,m})~\dots ~x(t_{N_m,m})]\]
where $\mathscr{X}_m$ $\in \mathbb{R}^{n\times N_m}$  is the snapshot matrix of actual system for operating region $\mathscr{R}_m$, $n$ is the number of states, $N_m$ is number of sampling interval in the operating region ($|\mathscr{R}_m| = N_m$). Note that this step is performed online and the trajectories are generated at the beginning of the operating region at time $t_{0,m}$.

\subsubsection{Step 3: projection matrix and reduced order model generation}
After generating the operating regions and the corresponding trajectories, one reduced model is created for each operating region. The reduced model is generated based on the trajectory of the system. The main idea is to investigate the system trajectories and create state cluster sets for each operating region. The states having similar dynamics based on the state trajectories are put into the same clusters. In this work, the agglomerative hierarchical clustering \cite{steinbach_comparison_2000} is used. We use the Euclidean distance between trajectories as the distance measure for states. The main reason to choose agglomerative hierarchical clustering is because of the capability to define the distance threshold between the clusters instead of predefining the number of cluster sets. The distance threshold is a tuning parameter for the accuracy of reduced model. There are three commonly used linkage methods present in agglomerative hierarchical clustering (e.g. single, average, complete linkage). In this work, we use the average linkage, and it considers the average distance between each point in one cluster to every point in other clusters and it is calculated as follows: 
\[
D(p,q) = \frac{1}{n_pn_q}\sum_{i = 1}^{n_p}\sum_{j = 1}^{n_q}d(x_{pi},x_{qj}) 
\]
and  
\[ 
d(x_{pi},x_{qj})^2 = \sum_{k=1}^n(x_{pik}-x_{qjk})^2 
\]
where $p$ and $q$ are two clusters, $i$ and $j$ are data points within the clusters, $d$ is the Euclidean distance between $i$ and $j$, $n_p,n_q$ are the size of the clusters of $p$ and $q$ respectively and $n$ is the dimension of state vector. 

Let us consider $\mathscr{C}^{(m)} = \{\mathscr{C}_1^{(m)},\mathscr{C}_2^{(m)}, \dots, \mathscr{C}_r^{(m)}\}$ be the collection of clusters for operating region ($\mathscr{R}_m$) after the hierarchical clustering and $r$ is the order of the resulted reduced model. The resulted clusters have following properties: i) $\mathscr{C}_i^{(m)} \cap \mathscr{C}_j^{(m)} = \Phi$ and ii) $\mathscr{C}_1^{(m)} \cup \mathscr{C}_2^{(m)} \cup\ldots \cup \mathscr{C}_r^{(m)} =N_x$, where $N_x$ is the total number of states. 

The adaptive reduced-order system is constructed based on the Petrov-Galerkin projection framework \cite{antoulas_approximation_2005}. For the Petrov-Galerkin projection method, the projection matrix is required. The projection matrix project the high dimensional system to a lower dimensional system. 
The projection matrix ($\mathscr{U}^{(m)} \in \mathbb{R}^{n\times r}$) is generated based on the state clusters ($\mathscr{C}^{(m)}$) and the elements of $\mathscr{U}^{(m)}$ are expressed as follows:
\[\begin{aligned}
\mathscr{U}^{(m)}_{i,j}=\left\{
\begin{array}{@{}ll@{}}
w_i, & \text{if}\ \text{vertex} \ i \in \mathscr{C}^{(m)}_j \\
0, & \text{otherwise}
\end{array}\right.
\end{aligned}
\]
and $\mathit{w_i}$ is determined as follows:
\[
\begin{aligned}
\mathit{w_{i}}&= 1/||\alpha_i|| \\
\alpha_i&= \mathbb{E}_i ^{T}\alpha
\end{aligned}
\]
where $\alpha = [1,\dots,1]^T \in \mathbb{R}^{n}$, $||\alpha_i||$ is the $L_2$ norm of $\alpha_i$, $\mathbb{E}_i = e_{\mathscr{C}_i}\in \mathbb{R}^{n\times m}$, $e_j$ is the $j$-th column of the identity matrix of size $\mathbb{R}^{n\times n}$ and $m$ is the cardinality of $\mathscr{C}^{(m)}_i$ set.

 The adaptive reduced model of (\ref{eq:nonlinear}) for each operating region ($\mathscr{R}_m \in \mathscr{R}$) is expressed as:
\begin{equation}
\begin{aligned}
\dot{\xi}^{(m)}(t)&= f_r^{(m)}(\xi^{(m)}(t), u(t)) 
\end{aligned}
\label{eq:red_nonlinear}
\end{equation}
where $f_{r}^{(m)}(\xi^{(m)}(t), u(t))= \mathscr{U}^{(m)T}f(\mathscr{U}^{(m)}\xi^{(m)}(t), u(t))$ and $\xi^{(m)}(t) = \mathscr{U}^{(m)T}x(t)$. Note that the actual state $x$ can be approximated based on mapping $\tilde{x}(t) = \mathscr{U}^{(m)}\xi^{(m)}$ in operating region ($\mathscr{R}_m$). The discrete model of (\ref{eq:red_nonlinear}) is expressed as follows: 
\begin{equation}
\begin{aligned}
\xi^{(m)}(k+1)&= f_{rd}^{(m)}(\xi^{(m)}(k), u(k)) 
\end{aligned}
\label{eq:red_nonlineard}
\end{equation}

\subsubsection{Step 4: information exchange between reduced models}
\begin{figure}[H]
\centerline{\includegraphics[width=0.7\columnwidth]{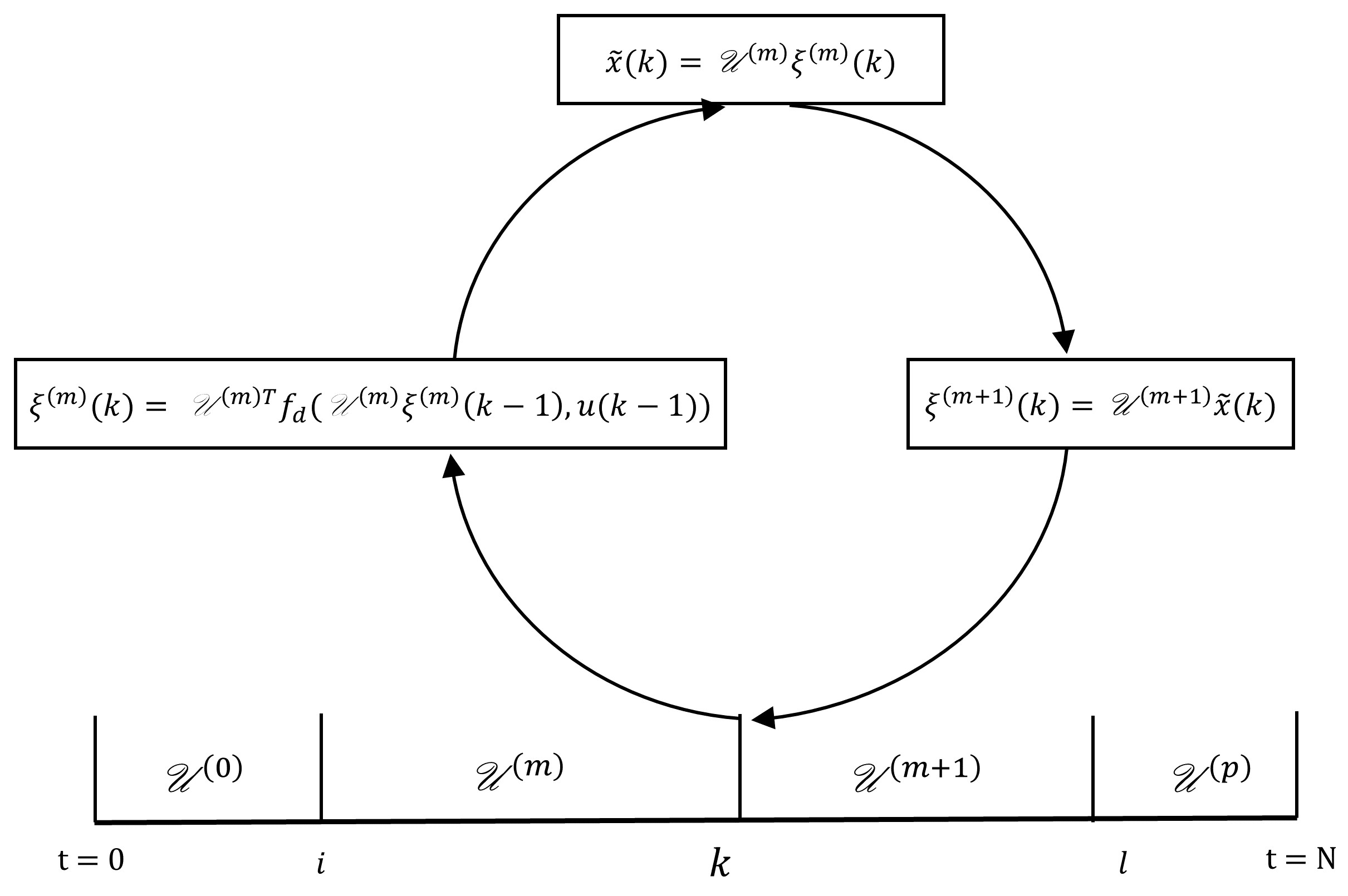}}
\caption{Changing boundary from one reduced model to another.} \label{fig:step_4}
\end{figure}

For each operating region, we may get a different order reduced-order model. Thus, when the system moves from one operating region to another operating region, the corresponding reduced models are also required to be switched. To handle the impact of changing reduced-order models from one operating region to other operating regions, we consider information exchange between reduced models at the boundary of the operating regions.
The steps to calculate the changing boundary is shown in Figure \ref{fig:step_4} and the algorithm as follows: 

    \begin{enumerate}
    \item Let us consider two operating regions $\mathscr{R}_m$ and $\mathscr{R}_{m+1}$ at boundary of time $k$.
    Compute the states of the reduced system at $k$ using the projection matrix $\mathscr{U}^{(m)}$ which is generated for the operating region $\mathscr{R}_m$ and, the past reduced state and input information. 
    \[\xi^{(m)}(k) = \mathscr{U}^{(m)T}f_{d}(\mathscr{U}^{(m)}\xi^{(m)}(k-1), u(k-1))\]
    \item Covert the reduced system to approximated original system state ($\tilde{x}(k)$) at $k$ using: $\tilde{x}(k)= \mathscr{U}^{(m)}\xi^{(m)}(k)$
    \item Compute the initial value for next operating region reduced model as $\xi^{(m+1)}(k) = \mathscr{U}^{(m+1)}\tilde{x}(k)$
    \end{enumerate}

\section{Adaptive Moving Horizon Estimation}
\label{section:estimation}
This section proposes the adaptive moving horizon estimation technique for the state estimation of the adaptive reduced-order model. As the reduced model order changes over time, using the classical moving horizon method is not feasible. So in this section, we propose the adaptive MHE design. 
In the proposed design, the length of the operating region is considered same length as the step size of the reduced order model. In other word, the reduced model order may change over each time step. The details of proposed adaptive MHE are described as follows. 

Let us consider the discrete reduced model (\ref{eq:red_nonlineard}) at each operating region with additive process and measurement noise as follows:
\begin{equation}
    \begin{aligned}
\xi^{(k)}(k+1)= \quad & f_{rd}^{(k)}(\xi^{(k)}(k),u(k)) + w_r^{(k)}(k) \\
y(k) = \quad & C\mathscr{U}^{(k)}\xi^{(k)}(k)+v(k)  
\end{aligned}
\label{eq:red_system_w}
\end{equation}
where $w_r^{(k)}(k)$ denotes the process noise of the reduced model at time $k$ and operating region $k$ and $v(t)$ is the measurement noise. Note that if the original nonlinear system also has additive process noise and is $w(k)$, then $w_r^{(k)}(k)=\mathscr{U}^{(k),T}w(k)$.  In MHE approach, for time ($k$) less than horizon length ($N$), the full information problem Equation (\ref{eq:full_info}) is solved while the time ($k$) greater than horizon length ($N$), the MHE problem with arrival cost is solved with Equation (\ref{eq:mhe_red}). 

\begin{subequations}
\begin{align}
  \phi_k= 
\min\limits_{\hat{\xi}^{(k)}_{0|k}, \{\hat{w_r}^{(k)}\}_{k=0}^{k-1}} &  
\sum\limits_{j=0}^{k-1}||\hat{w_r}^{(k)}(j)||^2_{\{Q_r^{(k)}\}^{-1}} + \sum\limits_{j=0}^{k}||\hat{v}(j)||^2_{R^{-1}} + ||\hat{\xi}^{(k)}_{0|k}-\bar{\xi}^{(k)}_0 ||^2_{\{P_r^{(k)}\}^{-1}}
\vspace{2mm}  \label{eq:full_info_a}         \\
{\rm s.t.~} &   \hat{\xi}^{(k)}(j+1) =  f_{rd}^{(k)}(\hat{\xi}^{(k)}(j),u(j)) + w_r^{(k)}(j) ,~~ j \epsilon [0, k-1] \vspace{2mm} \label{eq:full_info_b} \\
& \hat{v}(j) = y(j) - C_r^{(k)} \hat{\xi}^{(k)}(j), ~~ j \epsilon [0, k]\vspace{2mm} \label{eq:full_info_c} \\
& \hat{\xi}^{(k)}(j) \in \m E^{(k)}, ~ \hat{w_r}^{(k)}(j) \in \m W^{(k)}, ~ \hat{v}(j) \in \m V_f  \label{eq:full_info_d}  \vspace{2mm}
\end{align}
\label{eq:full_info}
\end{subequations}
\vspace{-10mm}
\begin{subequations}
\begin{align}  \Phi_k= 
\min\limits_{\hat{\xi}^{(k)}_{k-N|k}, \{\hat{w_r}^{(k)}\}_{k=k-N}^{k-1}} & 
\sum\limits_{j=k-N}^{k-1}||\hat{w_r}^{(k)}(j)||^2_{\{Q_r^{(k)}\}^{-1}} + \sum\limits_{j=k-N}^{k}||\hat{v}(j)||^2_{R^{-1}} + ||\hat{\xi}^{(k)}_{k-N|k}-\bar{\xi}^{(k)}_{k-N} ||^2_{\{P_r^{(k)}\}^{-1}}
\vspace{2mm} \label{eq:mhe_red_a} \\
{\rm s.t.~} &   \hat{\xi}^{(k)}(j+1) =  f_{rd}^{(k)}(\hat{\xi}^{(k)}(j),u(j)) + w_r^{(k)}(j) ,~~ j \epsilon [k-N, k-1] \vspace{2mm} \label{eq:mhe_red_b} \\
& \hat{v}(j) = y(j) - C_r^{(k)} \hat{\xi}^{(k)}(j), ~~ j \epsilon [k-N, k]\vspace{2mm} \label{eq:mhe_red_c} \\
& \hat{\xi}^{(k)}(j) \in \m E^{(k)}, ~ \hat{w_r}^{(k)}(j) \in \m W^{(k)}, ~ \hat{v}(j) \in \m V_f\label{eq:mhe_red_d}
\end{align}
\label{eq:mhe_red}%
\end{subequations}where $\hat{w}_r^{(k)}$ denotes the estimated reduced order system disturbance, $\hat{v}_r^{(k)}$ denotes the estimated measurement noise and $\hat{\xi}_r^{(k)}$ denotes the estimated reduced order system states. In Equations (\ref{eq:full_info_a},~\ref{eq:mhe_red_a}), the $Q_r^{(k)}$ and $R$ are symmetric positive definite weight matrices which penalize the estimated system noise and output noise respectively. The weighting matrix $P_r^{(k)}$ makes sure that the estimated system state ($\hat{\xi}^{(k)}_0$) is not far away from the initial guess of the system ($\bar{\xi}^{(k)}_0$). 
The $Q_r^{(k)}$ is calculated based on the projection of full order system $Q$ matrix ($Q_r^{(k)}= \mathscr{U}^{(k)}Q\mathscr{U}^{(k)T}$). Similarly the weighting matrix $P_r^{(k)}$ is calculated using full order system $P$ matrix ($P_r^{(k)}= \mathscr{U}^{(k)}P\mathscr{U}^{(k)T}$). The Equations (\ref{eq:full_info_b},~\ref{eq:mhe_red_b}) are the reduced order models with process noise and the Equations (\ref{eq:full_info_c},~\ref{eq:mhe_red_c}) are the output models with measurement noise. Equations (\ref{eq:full_info_d},~\ref{eq:mhe_red_d}) represent the constraints for the estimated system states, estimated process noise and estimated measurement noise.

Let us first discuss the steps for adaptive full information estimation problem. First we obtain the measurements ($y_{0:k}$) from time $0$ to current time $k$. Then the snapshot matrix is generated based on simulating the original non-linear system from time $t=0$ to $t=k+1$ using the initial guess $\bar{x}_0$ and input. After that the projection matrix ($\mathscr{U}^{(k)}$) is generated for current time $k$ as discussed in Section 3. The reduced model is generated as per Equation (\ref{eq:red_system_w}) using the generated projection matrix. The initial guess of the full order system states is converted into reduced order guess using $\bar{\xi}^{(k)}_0 = \mathscr{U}^{(k)T}\bar{x}_0$. Then the optimization problem Equation (\ref{eq:full_info}) is solved using the initial guess of the reduced order system and measurements collected from time zero. After solving the optimization problem, we obtain the current estimates of the reduced order system states $\hat{\xi}^{(k)}_{j|k}$ from time $j=0$ to $j=k$. At the final step, the estimated reduced order system state at current time is projected back into full order system states using ($\hat{\tilde{x}}_{k|k}= \mathscr{U}^{(k)}\hat{\xi}^{(k)}_{k|k}$) and then continue the loop till $k<N$. 

The moving horizon estimation at ($k>N$) also follows the similar algorithm as the full information estimation problem. First the measurement is collected from $t = k-N$ to current time $k$. Then snapshot matrix is generated from time $k-N$ to $k+1$ based on initial condition $\bar{x}_{k-N}$ which is the estimated state $\hat{\tilde{x}}_{k-N|k-N}$ obtained at $k-N$. Then the projection matrix followed by the reduced order non-linear model is calculated. In the next step, the reduced-order system initial guess for the optimization problem (\ref{eq:mhe_red}) is calculated using $\bar{\xi}^{(k)}_{k-N} = \mathscr{U}^{(k)T}\bar{x}_{k-N}$. After that the optimization problem Equation (\ref{eq:mhe_red}) is solved to obtain the reduced order estimates $\hat{\xi}^{(k)}_{j|k}$ for $j = k-N, \dots, k$. Next, we obtain the estimate of approximated full order system states using the back projection $\hat{\tilde{x}}_{k|k}= \mathscr{U}^{(k)}\hat{\xi}^{(k)}_{k|k}$. Finally  increase the time, update the initial guess and continue the loop till final time. The details of the adaptive MHE algorithm is presented in Algorithm 1.

\begin{algorithm}[t]
\caption{Adaptive moving horizon estimation algorithm}
\begin{algorithmic}[1]

\INPUT {Initial guess $\bar{x}_0$, covariance matrices ($P_0>0, Q, R>0$), horizon length ($N\geq1$), simulation length ($T$)}
\STATE \textbf{Initialization} $\bar{x}_{k-N} \leftarrow \bar{x}_0$
\OUTPUT Estimated states in the order of actual system
\FOR{$k = 0 \dots T$}
\IF{$k < N$}
\STATE Obtain measurements $Y = y_{0:k}$
\STATE Generate snapshot matrix : $X = [\bar{x}_0 \dots ~\bar{x}_{k+1}]$
\STATE Generate projection matrix ($\mathscr{U}^{(k)}$)
\STATE Obtain reduced nonlinear model using Equation~(6)
\STATE Covert the initial guess to reduced order guess $\bar{\xi}^{(k)}_0 = \mathscr{U}^{(k)T}\bar{x}_0$
\STATE Solve $\phi_k$ with reduced nonlinear model, $\bar{\xi}^{(k)}_0$ and $Y$
\STATE Obtain current estimates $\hat{\xi}^{(k)}_{j|k}$ for $j = 0, \dots, k$ 
\STATE Convert the reduced order estimate to approximated original state $\hat{\tilde{x}}_{k|k}= \mathscr{U}^{(k)}\hat{\xi}^{(k)}_{k|k}$
\ELSE
\STATE Obtain measurements $Y = y_{k-N:k}$
\STATE Generate snapshot matrix : $X = [\bar{x}_{k-N} \dots ~\bar{x}_{k+1}]$ with $\bar{x}_{k-N}$ as initial condition
\STATE Generate projection matrix ($\mathscr{U}^{(k)}$)
\STATE Obtain reduced nonlinear model using Equation~(6)
\STATE Covert the initial guess to reduced order guess $\bar{\xi}^{(k)}_{k-N} = \mathscr{U}^{(k)T}\bar{x}_{k-N}$
\STATE Solve $\Phi_k$ with reduced nonlinear model, $\bar{\xi}^{(k)}_{k-N}$ and $Y$
\STATE Obtain current reduced order estimates $\hat{\xi}^{(k)}_{j|k}$ for $j = k-N, \dots, k$ 
\STATE Convert the estimates to approximated original states $\hat{\tilde{x}}_{k|k}= \mathscr{U}^{(k)}\hat{\xi}^{(k)}_{k|k}$ 
\STATE $\bar{x}_{k-N} \leftarrow \hat{\tilde{x}}_{k-N+1|k-N+1}$
\ENDIF
\ENDFOR
\end{algorithmic}
\end{algorithm}

\section{Application to a small field: simulation case}
In this section, we apply the proposed algorithms to a small simulated field. First, the performance of the proposed adaptive MHE is compared with the original MHE results. The original MHE is designed based on the full-order system. Secondly, the robustness of the adaptive MHE is discussed based on the different initial guesses of the estimator.
The main motive to use the small simulated field in this section is that it is convenient to simulate the original MHE for a small-scale system. As the number of nodes increases, the computational cost increases exponentially for the original MHE. The original MHE used in this simulation is formulated as follows: 
\vspace{3mm}
\begin{subequations}
\begin{align}  \Gamma_k= 
\min\limits_{\hat{x}_{k-N}, \{\hat{w_k}\}_{k=k-N}^{k-1}} &  
\sum\limits_{j=k-N}^{k-1}||\hat{w}(j)||^2_{Q^{-1}} + \sum\limits_{j=k-N}^{k}||\hat{v}(j)||^2_{R^{-1}} + ||\hat{x}_{k-N}-\bar{x}_{k-N|k-N} ||^2_{P^{-1}}
\vspace{2mm}\label{eq:or_mhe_a}\\
{\rm s.t.~} &   \hat{x}(j+1) =  f(\hat{x}(j),u(j))+w(j)   ,~~ j \epsilon [k-N, k-1] \vspace{2mm} \label{eq:or_mhe_b}\\
& \hat{v}(j) = y(j) - C \hat{x}(j), ~~ j \epsilon [k-N, k]\vspace{2mm} \label{eq:or_mhe_c}\\
& \m X_{lb} \leq\hat{x}(j) \leq \m X_{ub}, ~ \m W_{lb} \leq \hat{w}(j) \leq \m W_{ub}, ~ \hat{v}(j) \in \m V \label{eq:or_mhe_d}
\end{align}
\label{eq:or_mhe}%
\end{subequations}
where $\hat{x}_{K-N}$ and $\hat{w}_k$ are the decision variables to the optimization problem. $N$
denotes the length of the optimization problem, and $k$ is the current time instance. The optimization problem objective is to minimize the error of predicted and measurements ($||\hat{v}||^2_{R^{-1}}$), the process disturbance ($||\hat{w}||^2_{Q^{-1}}$) and the arrival cost ($|.|^2_{P^{-1}}$). The arrival cost summarizes the cost from the initial time to the model until the beginning of the estimation window. The states and model uncertainty is bounded. $P, Q, R$ are the constant positive define matrices and the tuning parameters of the optimization problem. $P$ matrix is calculated based on the following formula: 
\begin{equation}
    P = l \times M^TM
    \label{eq:P}
\end{equation}
where $M = 0.5(\m X_{lb} - \m X_{ub})$ and $l$ is a tuning parameter.

\begin{figure}[t]
\centering
\includegraphics[width=0.9\columnwidth]{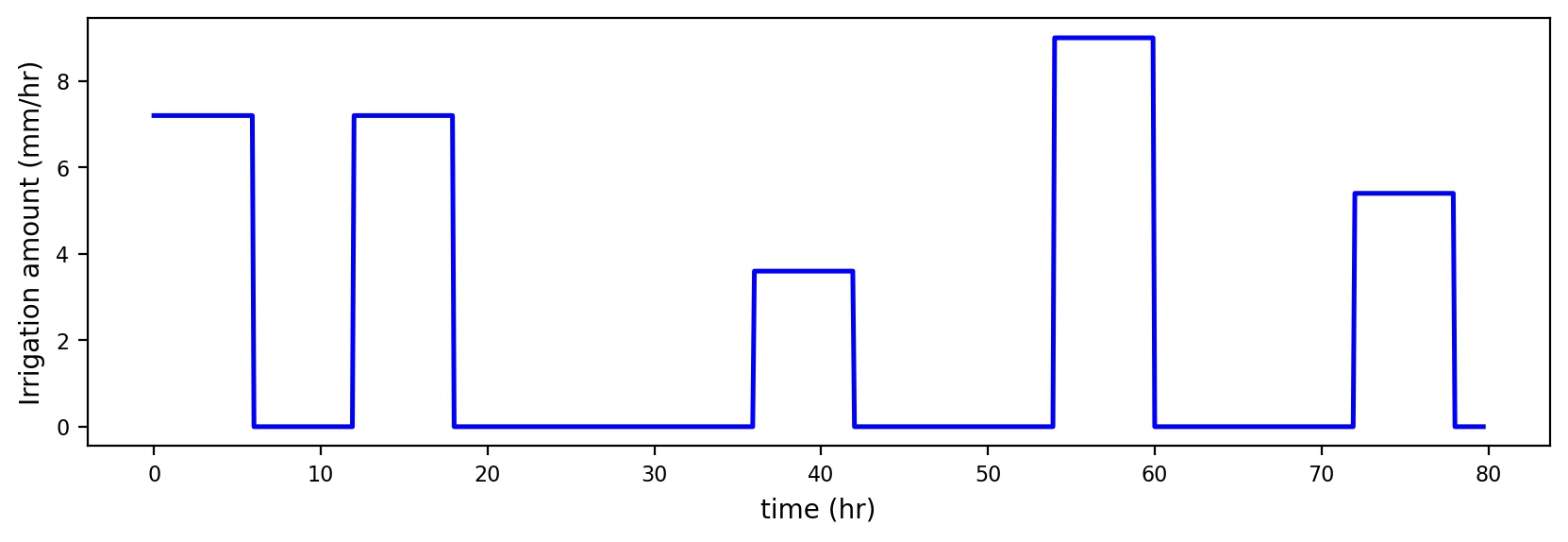}
\caption{Irrigation amount for the small field}
\label{fig:input}
\end{figure}

\begin{table}[t]
	\caption{The parameters of loamy soil.}
	\small 
	\centering
	\begin{tabular}{cccccc}
		\hline
		  {$K_{s}$ (m/s)} &{$\theta_{s}$ $(\text{m}^{3}/\text{m}^{3})$}&{$\theta_{r}$ $(\text{m}^{3}/\text{m}^{3})$}& {$\alpha$ (1/m)}& {$n$ (-)}\\
		\hline
		 $2.889\times 10^{-6}$  & 0.430 & 0.0780&3.60&1.56\\
		\hline
	\end{tabular} \label{tbl:soil_prop}
\end{table}
We consider a circular field of radius 10 m and depth 0.15 m with loam soil type. Table \ref{tbl:soil_prop} shows the hydraulic properties of the loam soil used in the simulation. 
The field is discretized into 400 nodes with 5 nodes in radial, 8 nodes in azimuthal and 10 nodes in the axial direction. The soil pressure heads at these discretization nodes are the states of the system. Total simulation time is 80hrs, and the time step used for temporal discretization is 6 mins. The central pivot is considered as the irrigation system, and the central pivot's rotating speed is 0.01745 m/s. Non-uniform irrigation has been considered for over 100 hours, and the irrigation amount is shown in Figure \ref{fig:input}. Sugar beet at its development stage is considered as the crop for the field. The reference evapotranspiration value is considered as 0.085 mm/hr, and the crop-coefficient value is 0.35. We consider 3 sensors at the depth of 13.5 cm (states 89, 185, 281). Process and measurement noise are considered for the simulation with zero mean and standard deviation of $0.15$ m and $0.15$ m, respectively. 
\subsection{Performance comparison of proposed adaptive MHE and original MHE}
In this subsection, we compare the performance of the proposed adaptive MHE with the original MHE. The actual system is used for generating measurements and it is simulated with a homogeneous initial condition of $-1.0$ m pressure head. 
The initial guesses for both the adaptive MHE and original MHE is considered the $50\%$ of the actual initial state. The estimation window size is 3. For the original MHE, the Q and R values are considered a diagonal matrix with $10^{-6}$ and $0.01$ values, respectively and P matrix is designed based on Equation (\ref{eq:P}). For the adaptive MHE, $Q_r, P_r$ in Equations (\ref{eq:full_info_a}, \ref{eq:mhe_red_a}) are calculated based on ($P_r^{(k)}= \mathscr{U}^{(k)}P\mathscr{U}^{(k)T}$) and ($Q_r^{(k)}= \mathscr{U}^{(k)}Q\mathscr{U}^{(k)T}$).

\begin{figure}[t]
\centerline{\includegraphics[width=1\columnwidth]{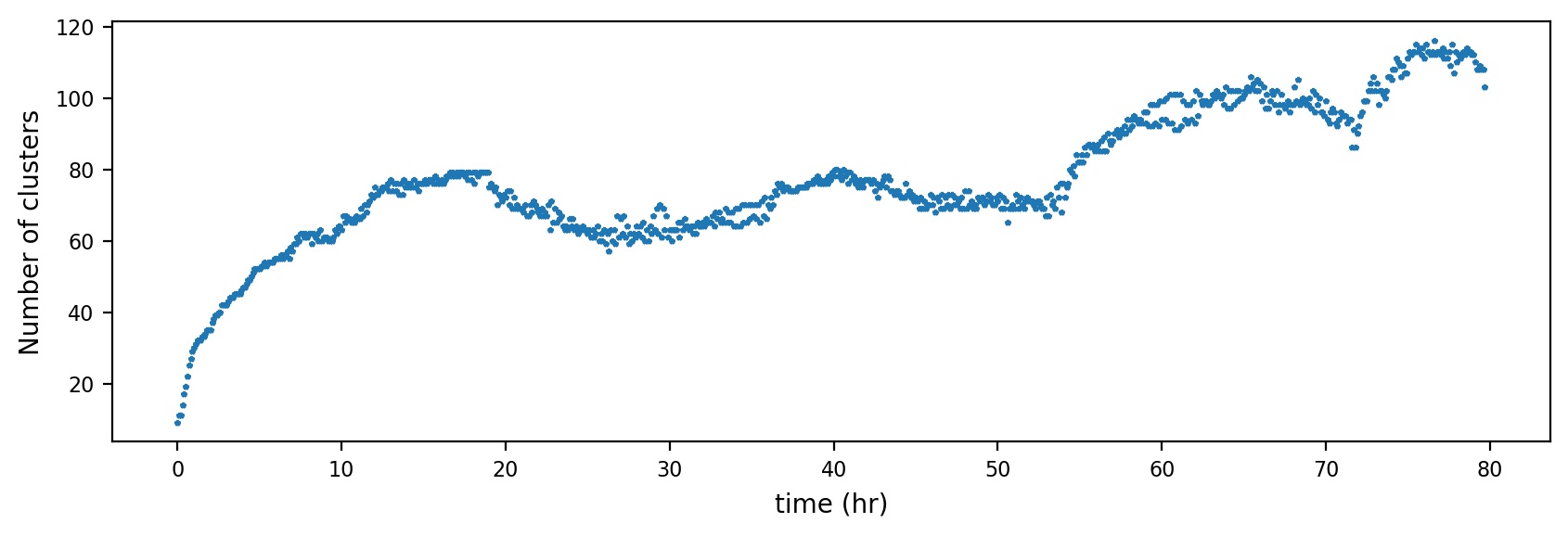}}
\caption{Number of clusters for the small field} 
\label{fig:Nc_case_s}
\end{figure}

\begin{figure}[t]
\centerline{\includegraphics[width=1\columnwidth]{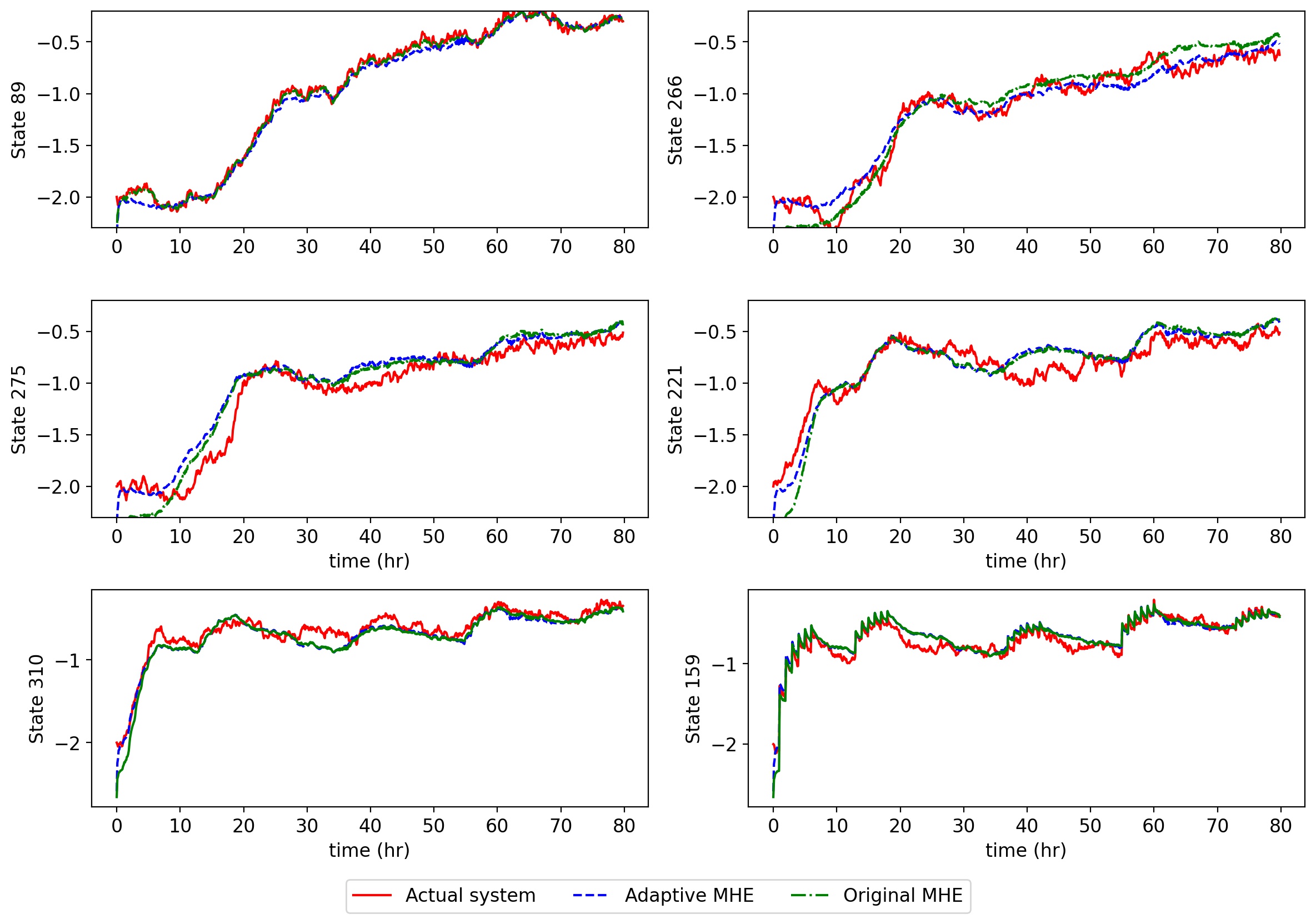}}
\caption{Selected state trajectories of the actual states (red solid line), estimated states using adaptive MHE (blue dotted line) and estimated states using original MHE (green dashed line)} \label{fig:Traj_case_s}
\end{figure}

First, the number of clusters each time for the adaptive MHE is shown in Figure \ref{fig:Nc_case_s}. It shows that the number of clusters decrease and increase based upon the dynamics of the system. Figure \ref{fig:Traj_case_s} shows some of the selected state trajectories of the actual system and the estimated states by the original MHE and the proposed adaptive MHE. It can be observed that both the original MHE and the proposed adaptive MHE can track the actual process very well.

\begin{figure}[t]
\centerline{\includegraphics[width=1\columnwidth]{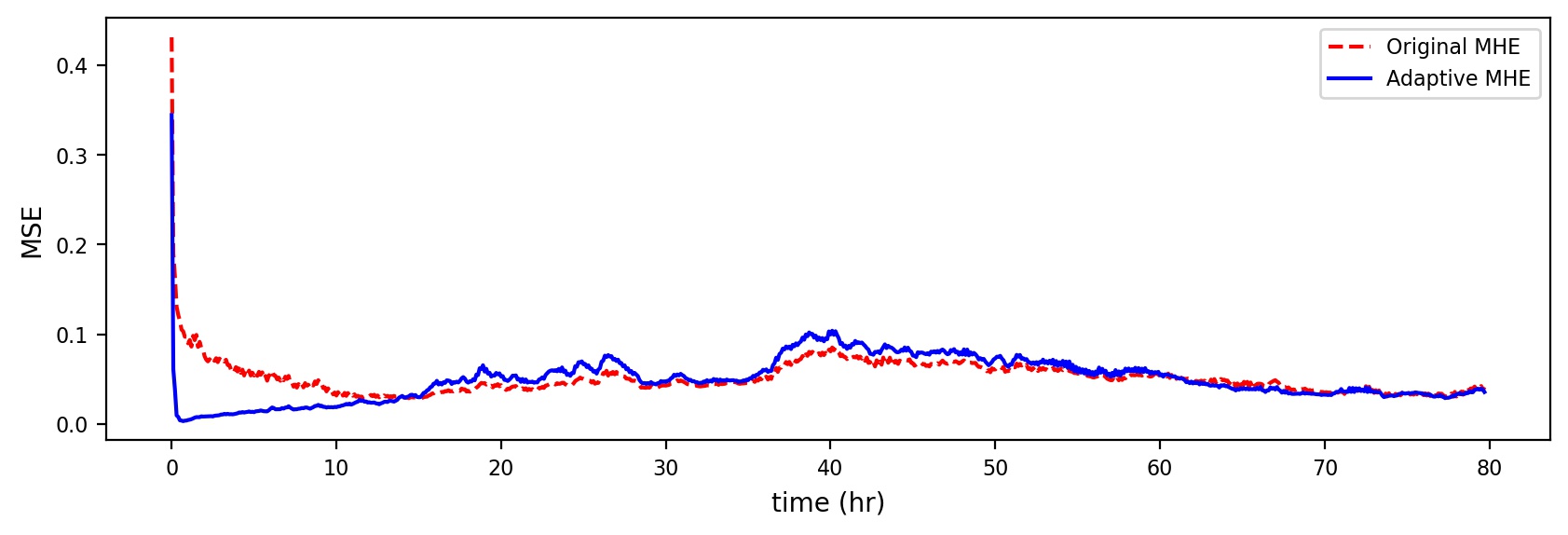}}
\caption{Mean square error of the original MHE (red solid line), adaptive MHE (blue solid line) and actual system with out noise (green solid line))} 
\label{fig:MSE_case_s}
\end{figure}

To further investigate the estimation performance, we consider the mean square error (MSE) of the estimation error as the performance indicator.
Figure \ref{fig:MSE_case_s} shows the MSE of the original MHE and adaptive MHE. It can be observed that the MSE of the original MHE is higher than the MSE of adaptive MHE at the beginning, and then it converges to the nearly identical value. It can be explained based on the degree of observability. From Figure \ref{fig:Nc_case_s}, we see that the number of clusters at the initial time is 9, which means the number of reduced states is 9. We have 3 measurements, the degree of observability to estimate 9 states in adaptive MHE is higher than that of estimating 400 states in original MHE. Therefore, in the beginning, the MSE of the adaptive MHE is lower than that of the original MHE. 
Another observation is that the system does not decrease to zero values. It is because of presence of noise in the process and measurement variables.

The computation speed of the original MHE and adaptive MHE is shown in Table \ref{tbl:speed}. It can be observed that the proposed adaptive speed is nearly 14 times faster than the original MHE.

\renewcommand{\arraystretch}{1.6}
\begin{table}[t]
\caption{Computational speed comparison of original MHE and adaptive MHE.}
	\centering
\begin{tabular}{ |c|c| } 
 \hline
 Original MHE & Adaptive MHE  \\ 
 \hline
 7773 sec & 546 sec \\ 
 \hline
\end{tabular}
\label{tbl:speed}
\end{table}

\subsection{Robustness of adaptive MHE}
\begin{figure}[t]
\centerline{\includegraphics[width=1\columnwidth]{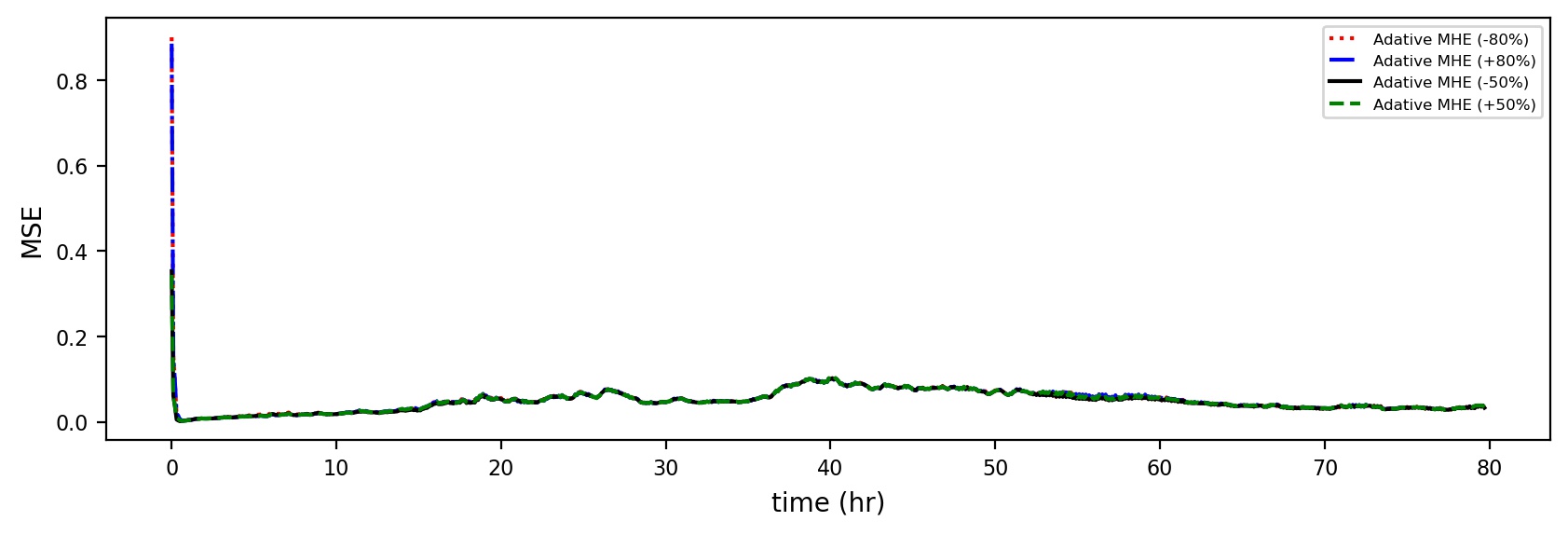}}
\caption{Mean square error of adaptive MHE staring from different initial guesses} 
\label{fig:Robust_case_s}
\end{figure}

More simulations are carried out to check the robustness of the proposed adaptive MHE. Different initial guesses to the adaptive MHE are provided in this simulation, varying from $+80\%, +50\%$, $-50\%, -80\%$ of the actual initial condition. Figure \ref{fig:Robust_case_s} shows the MSE of all 4 cases. We can observe that at the beginning, the $+80\%, -80\%$ cases have more error than the  $+50\%, -50\%$ cases, and after few time steps, all the cases converge to nearly same values. This result shows the robustness of the adaptive MHE. 

\section{Application to real-agricultural field}
In this section, the proposed procedure and algorithms are applied to perform model reduction and state estimation to a real-agricultural field. The agricultural field is located east of the city Lethbridge, Alberta, Canada (Lon: $-112.7385:-112.7365$, Lat: $49.6896: 49.6908$). The field is equipped with a central pivot irrigation system and consists of one pivot arm with 21 sprinklers. The radius of the field is 50m. One weather station is located near the agricultural field managed by Lethbridge Demo Farm Irrigation Management Climate Information Network (IMCIN). The weather station's data is obtained from the Alberta Climate Information Service website (https://agriculture.alberta.ca/acis/). For this work, the data from July $15^{th}$, 2019 is considered and at that period, the crop was sugar beet at its development stage. The crop coefficient ($K_c$) value 0.35 is used, and the reference evapotranspiration is obtained from the weather station to calculate the total evapotranspiration by using Equation (\ref{eq:ET}). The central pivot usually takes 8 hours to irrigate the whole field. 

\begin{figure}[t]%
\centering
\subfigure[]{
\includegraphics[width=0.35\columnwidth]{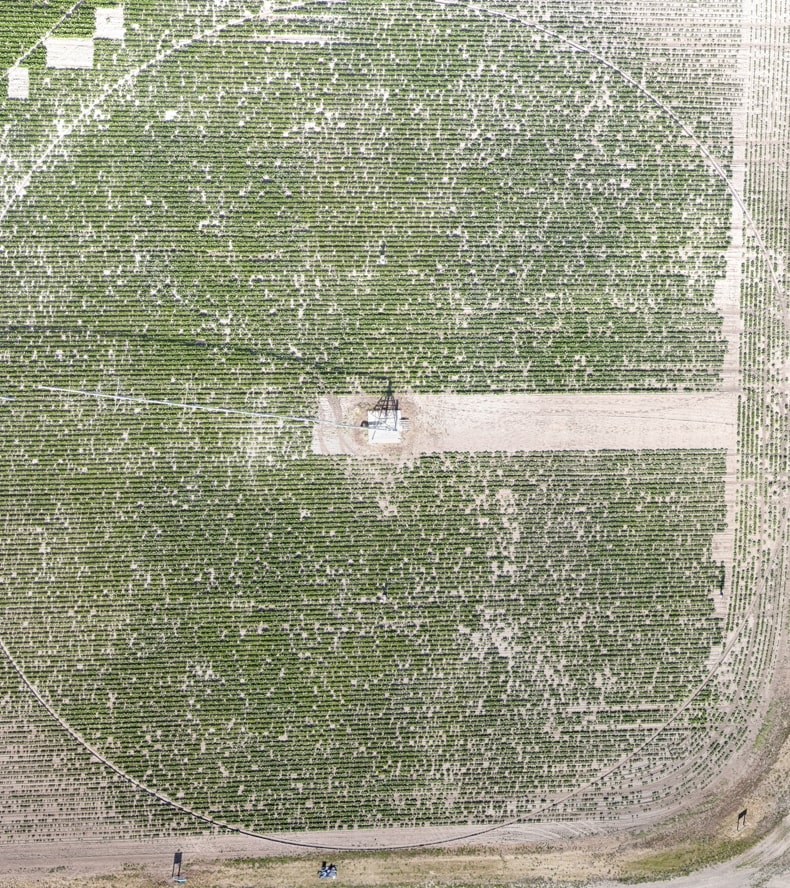}}
\qquad
\subfigure[]{
\includegraphics[width=0.55\columnwidth]{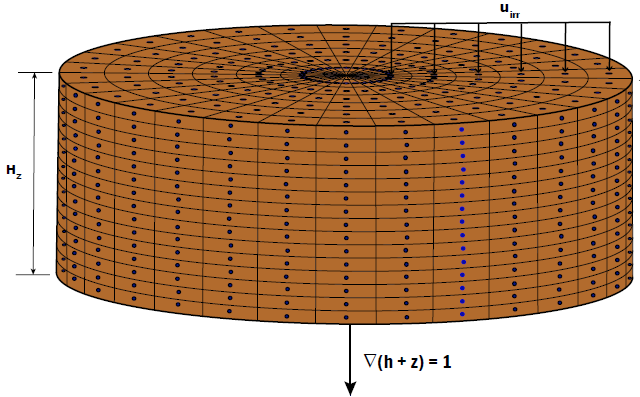}}
\caption{a) Demo farm in Lethbridge, (b) A schematic diagram of the demo farm model}
\label{fig:demofarm}
\end{figure}

In this work, the agro-hydrological model of the farm is constructed using finite difference discretization of Richards equation as discussed in section 2. For the model, the depth till 0.5 m is considered. The model is discretized into 31500 nodes with 21, 60, 25 nodes in the radial, azimuthal, and axial directions, respectively.
The demo farm and the schematic diagram of the demo farm is shown in Figure \ref{fig:demofarm}. 

\begin{figure}[t]
\centering
\subfigure[]{
\includegraphics[width=1\columnwidth]{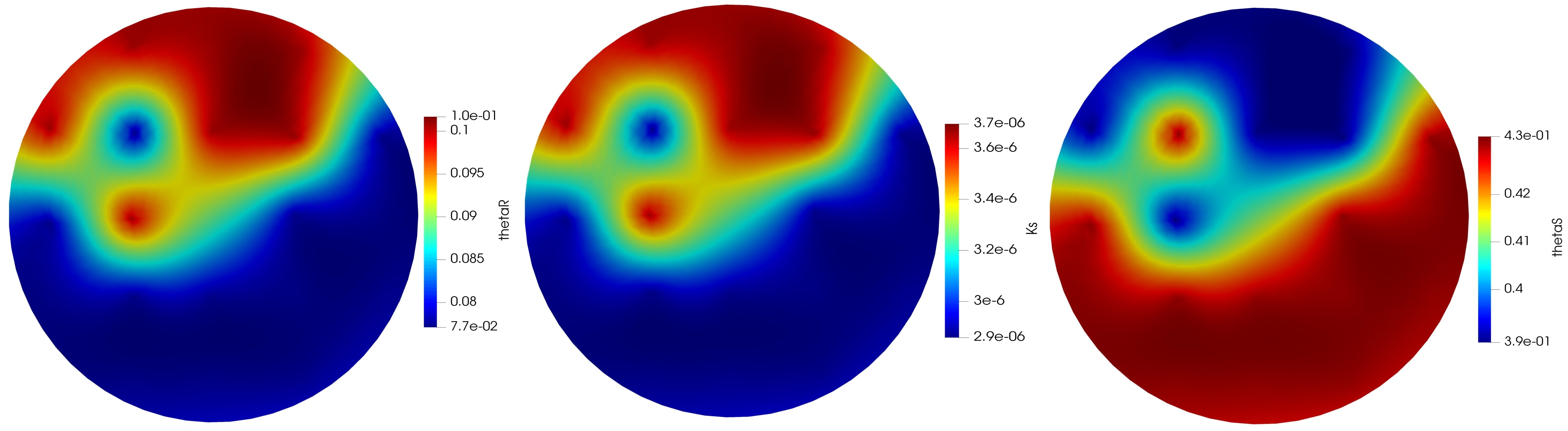}}
\qquad
\subfigure[]{
\includegraphics[width=1\columnwidth]{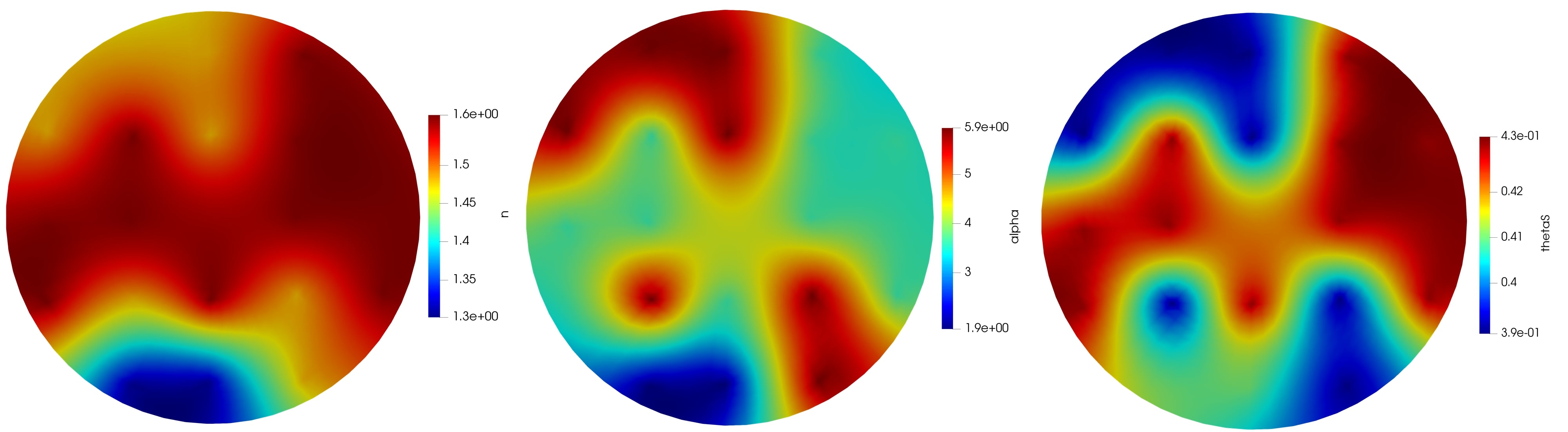}}
\caption{(a) soil parameters ($\theta_R, K_S, \theta_S$) from top to 25 cm depth, (b) soil parameters ($n, \alpha, \theta_S$) from 25 cm to 50 cm depth}
\label{fig:kriging}
\end{figure}

The soil parameters of the model are obtained from the soil texture experiment conducted in summer 2019. The five parameters value ($K_s, \alpha, n, \theta_s, \theta_r$) in Richards equation (\ref{eq:richards}) depend upon the types of soil. The soil samples are collected at 40 points in the field (20 points from surface to depth 0.25 cm, 20 points at 0.25 cm to 0.5 cm). After collecting the soil sample, the types of soil are estimated in the soil lab. The five parameters of the soil are calculated based on types of soil \cite{carsel_developing_1988}. However, there are soil samples at only 40 points in the field, the soil parameters for the other nodes are unknown. The traditional kriging approach is used to map all the parameters for the whole field.  For the initial analysis, the spherical kriging has been used to map all the parameters for the whole field. Figure \ref{fig:kriging} shows some of the selected parameters of the entire Demo farm after kriging interpolation. Note that this parameter estimation by kriging interpolation is not accurate enough. The accurate parameter estimation can be performed using remote sensing images and soil moisture probes using optimization approaches. The parameter estimation does not come under the scope of this work. 

\begin{figure}[t]
\centering
\includegraphics[width=0.4\columnwidth]{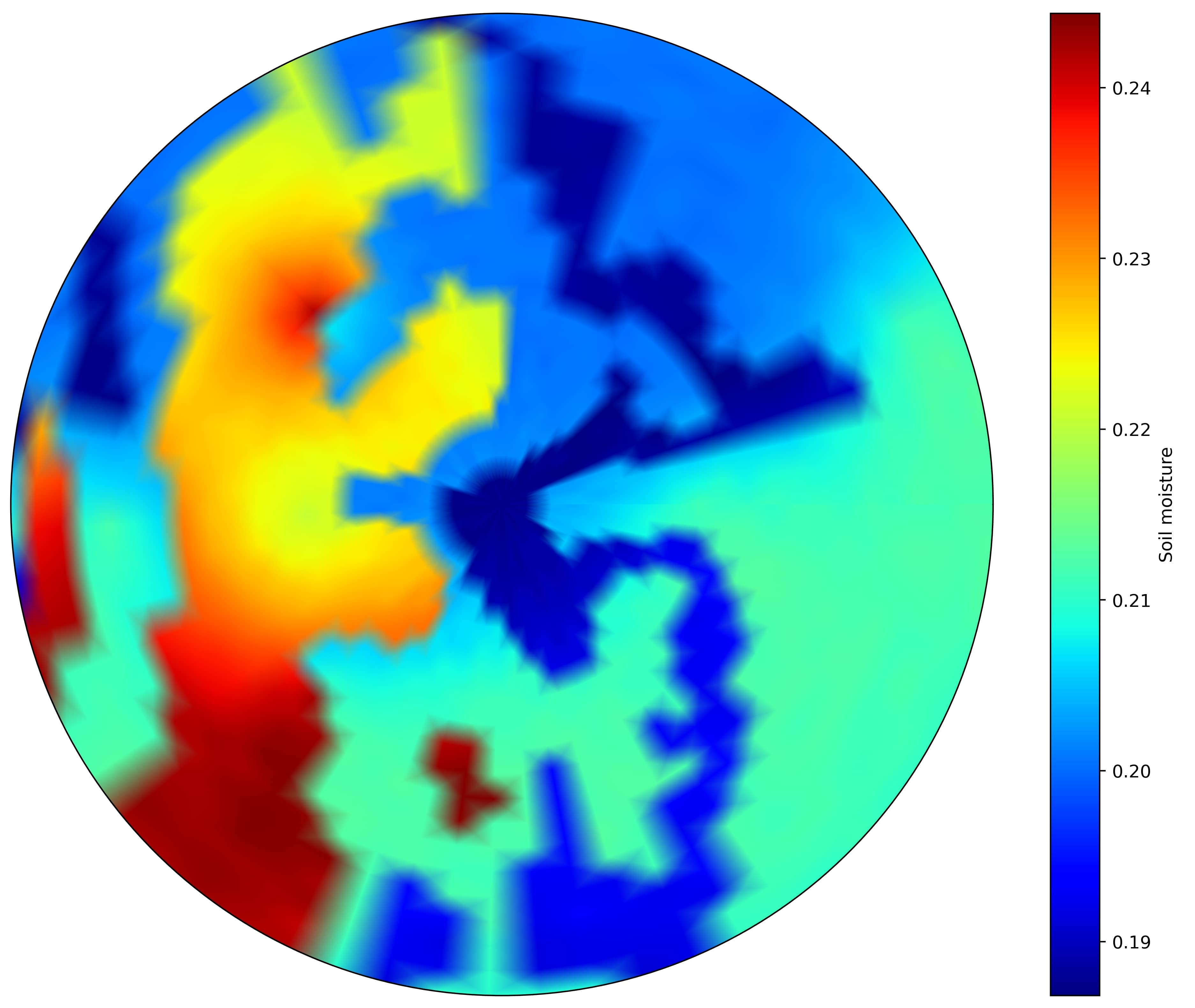}
\caption{Surface soil moisture initial condition}
\label{fig:surface_SM}
\end{figure}

The initial condition of the field is obtained from the thermal and optical remote sensing images. First, the surface soil moisture is estimated from the thermal and Normalized difference vegetation index (NDVI) data. The k-means clustering approach is applied to find out relative wet and dry areas of the field, and the clustered soil moisture is used as an initial condition of the demo farm. In this work, it is assumed the initial condition of the surface is the same in the z-direction till 50cm. Figure \ref{fig:surface_SM} shows the clustered surface soil moisture. We consider 20 sensors at the depth of 48 cm and the position of the nodes are 1526,  3101,  4676,  6251,  7826,  9401, 10976, 12551, 14126, 15701, 17276, 18851, 20426, 22001, 23576, 25151, 26726, 28301, 29876, 31451.

In the next subsections, the proposed procedure and algorithms are implemented on the real field model to perform adaptive model reduction and state estimation based on adaptive MHE under different scenarios. 

\subsection{Results: adaptive model reduction}
In this subsection, the proposed adaptive model reduction algorithm developed in Section \ref{section3} is applied to the real field. The main propose is to verify performance of the adaptive reduced order model for the large field. 

The initial condition, soil types and size of the field is considered as described above. For the simulation, the process noise is added. The process noise is normal distributed with zero mean and standard deviation of $6 \times 10^{-3}$ m. The irrigation amount is 0.72 mm/hr, and the total simulation time is 24 hr. 

\begin{figure}[t]
\centering
\includegraphics[width=1\columnwidth]{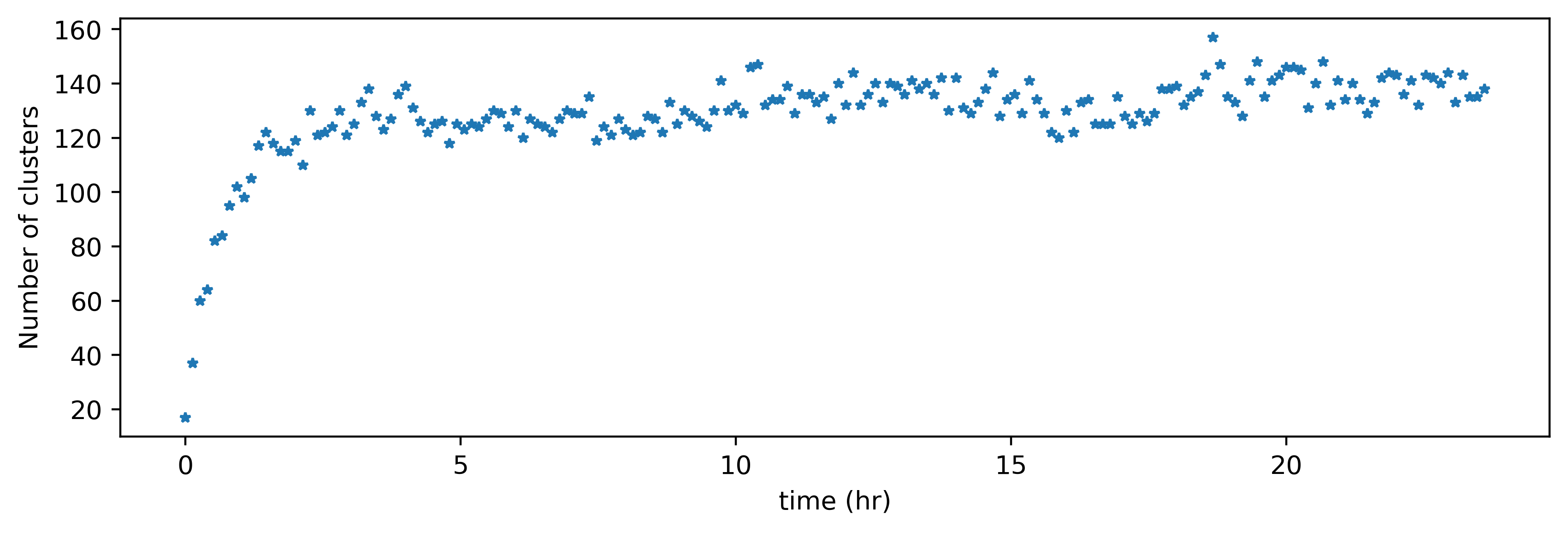}
\caption{Number of clusters for adaptive model reduction}
\label{fig:Nc_case0}
\end{figure}

\begin{figure}[t]
\centering
\includegraphics[width=1\columnwidth]{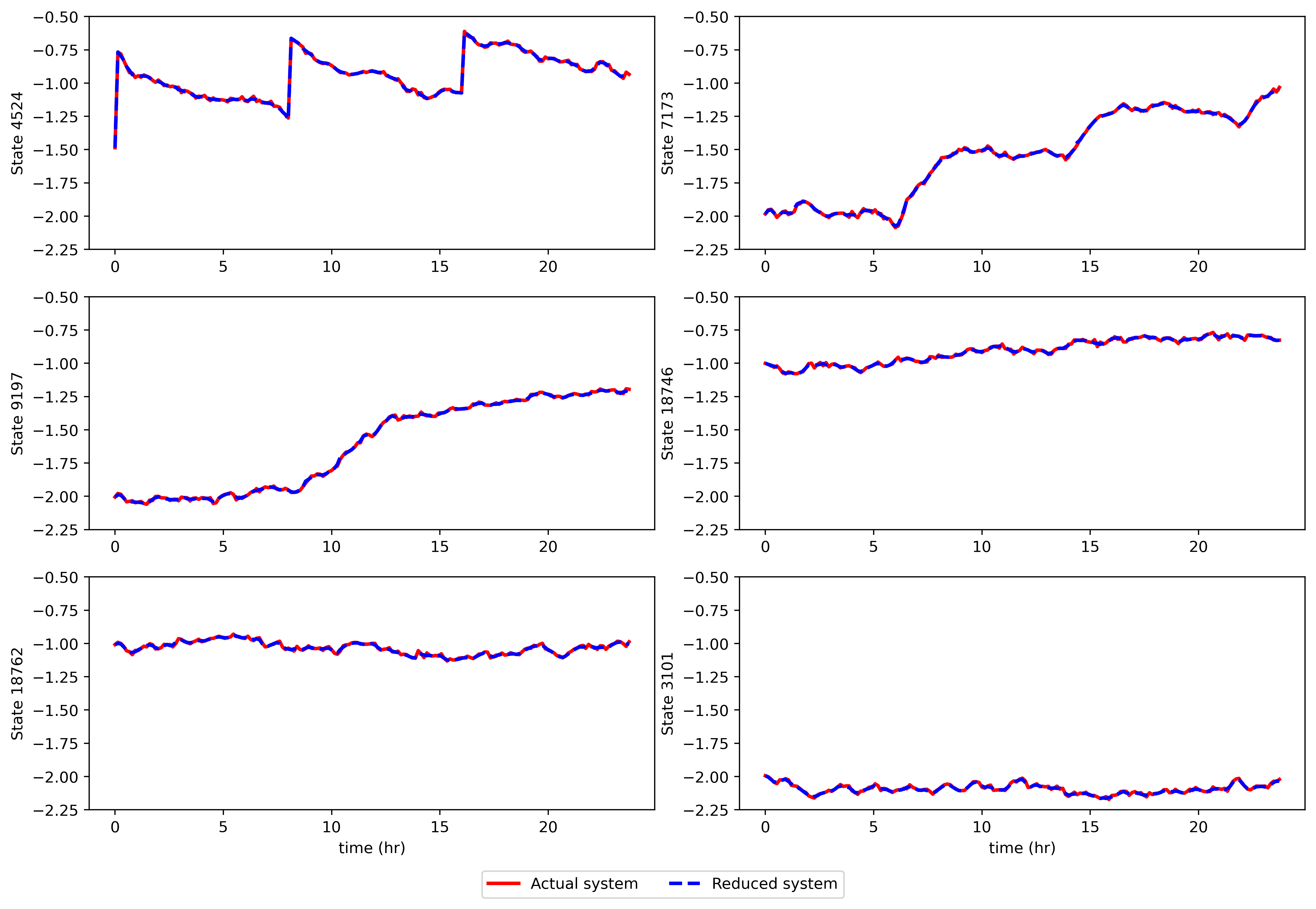}
\caption{Selected state trajectories of the actual states (red solid line), reduced states (blue dotted line) }
\label{fig:compare_traj_case0}
\end{figure}

\begin{figure}[t]
\centering
\includegraphics[width=1\columnwidth]{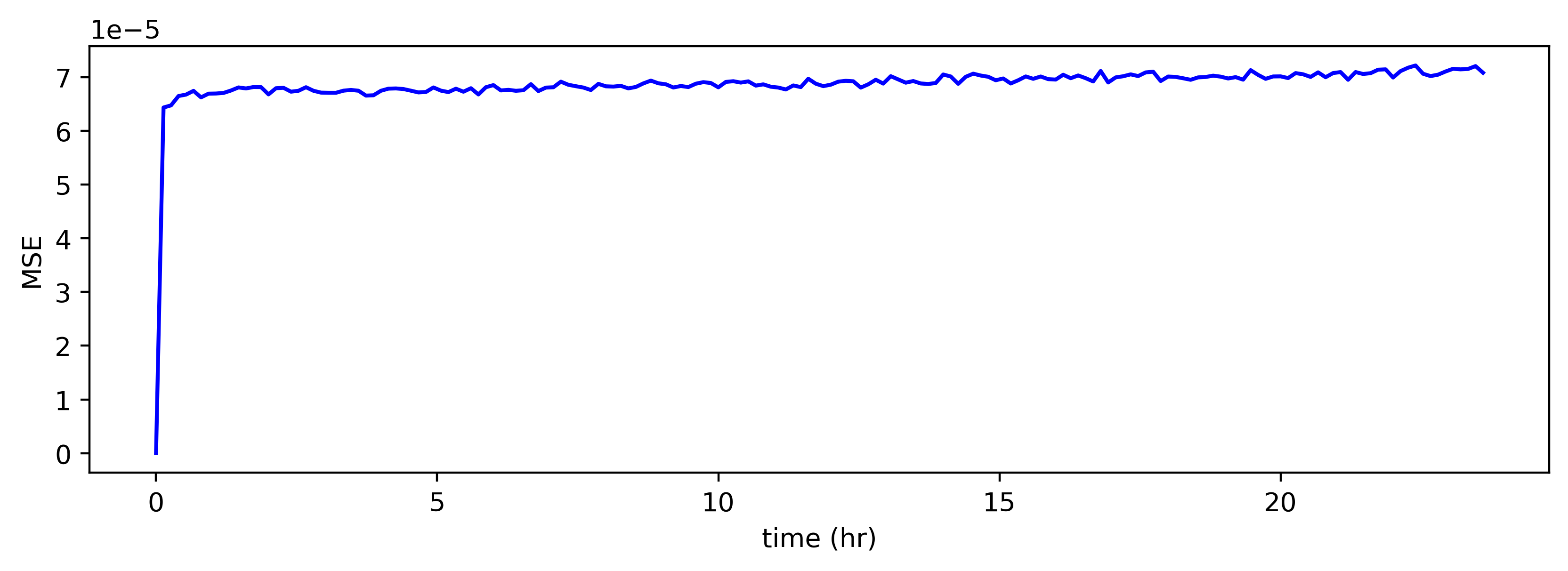}
\caption{Trajectory mean square error of actual system and reduced system}
\label{fig:MSE_case0}
\end{figure}

Figure \ref{fig:Nc_case0} shows the number of clusters each time after model reduction. It can be observed that based on the system dynamics, the number of clusters change each time.  Figure \ref{fig:compare_traj_case0} shows the state trajectories of the actual system and the adaptive reduced-order system. We can observe that the state trajectories of the actual system have an excellent agreement with the actual system trajectories. For further analysis of the performance, the MSE of the estimation error is considered and shown in Figure \ref{fig:MSE_case0}. It is evident that the MSE value is very low in the range of 1e-05, and thus the adaptive model reduction can track the actual system's trajectories accurately in presence of the process noise. 

\subsection{Result: adaptive MHE}
In this section, we apply the proposed adaptive MHE algorithms to the real field. The performance of the proposed adaptive MHE is demonstrated under the following two scenarios: (1) Scenario 1: field model without process and measurement noise; (2) Scenario 2: field model in presence of process and measurement noise.

\subsubsection{Scenario 1: nominal case} 
\begin{figure}[H]
\centering
\includegraphics[width=1\columnwidth]{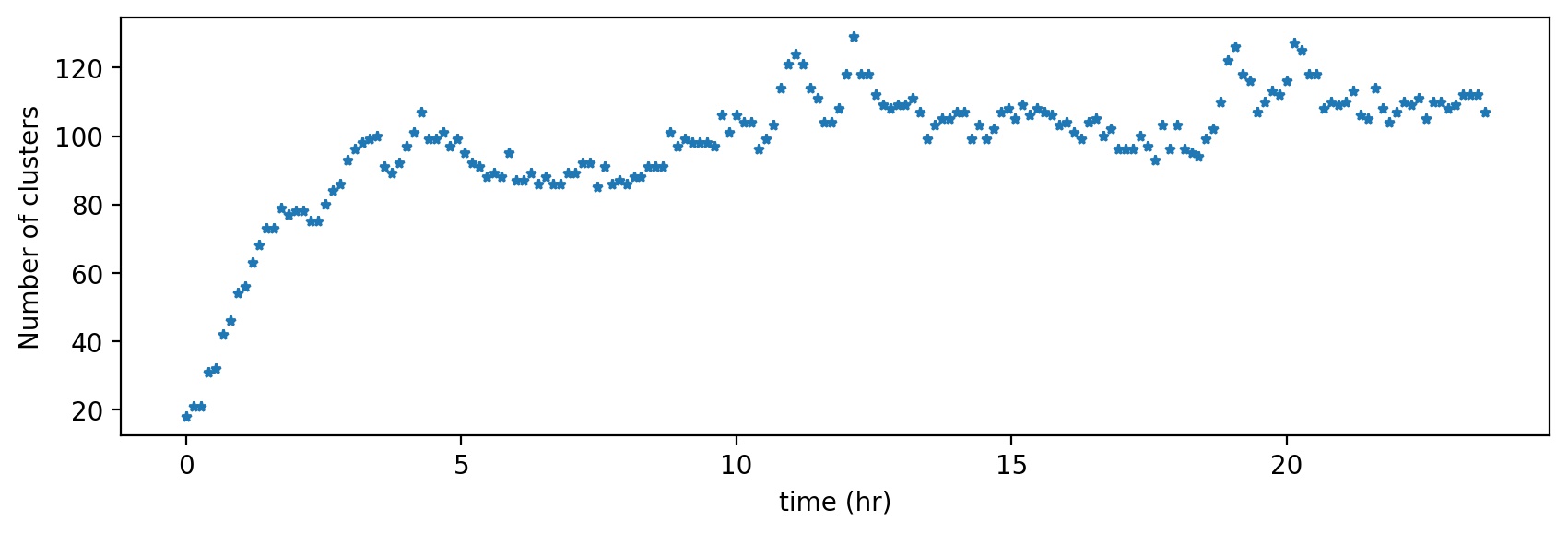}
\caption{Number of clusters for scenario 1}
\label{fig:NC_case1}
\end{figure}

\begin{figure}[h]
\centering
\includegraphics[width=1\columnwidth]{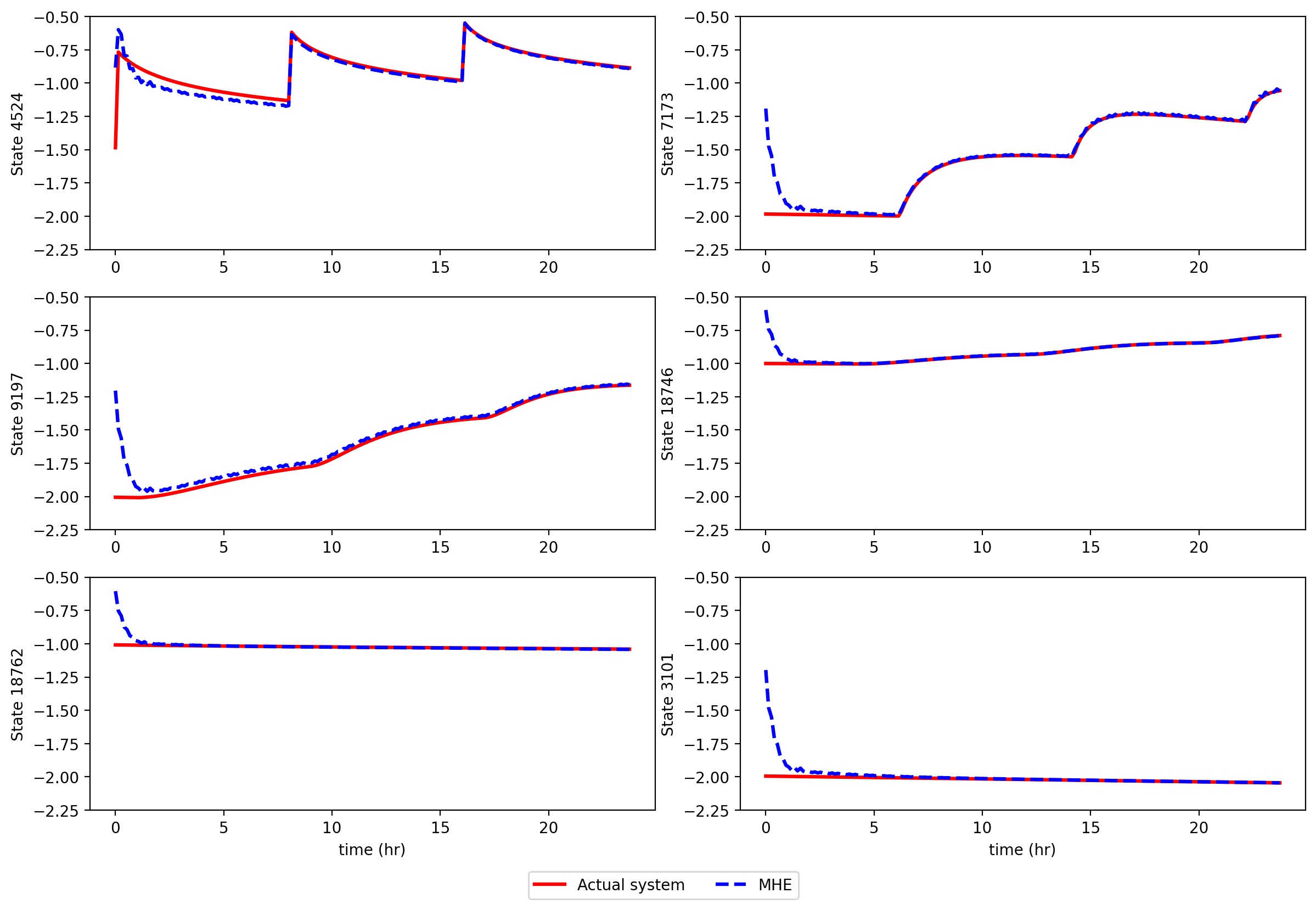}
\caption{Selected state trajectories of the actual states (red solid line), estimated states (blue dotted line) for scenario 1}
\label{fig:compare_traj_case1}
\end{figure}

\begin{figure}[H]
\centering
\includegraphics[width=0.8\columnwidth]{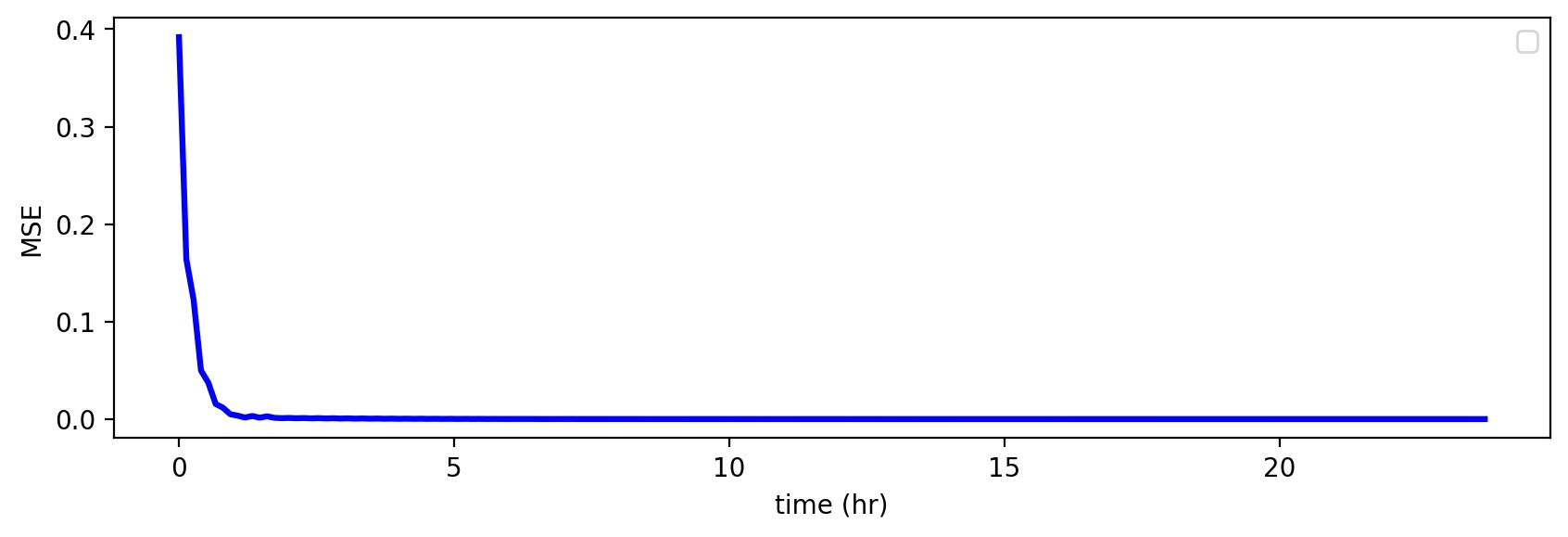}
\caption{Trajectory mean square error of adaptive MHE for scenario 1}
\label{fig:MSE_case1}
\end{figure}

\begin{figure}[H]
\centering
\subfigure[]{
\includegraphics[width=1\columnwidth]{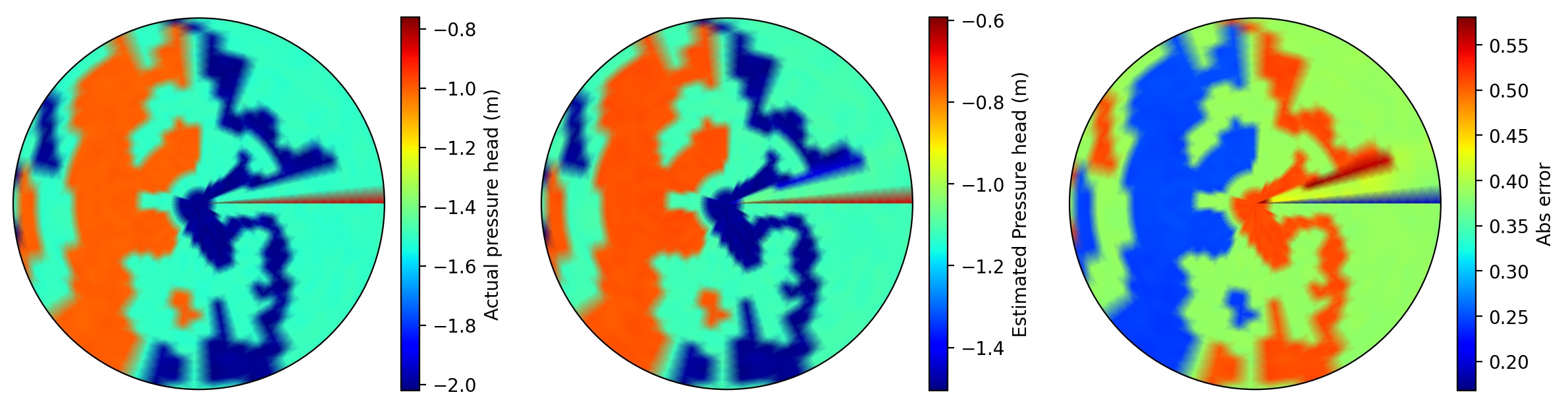}}
\qquad
\subfigure[]{
\includegraphics[width=1\columnwidth]{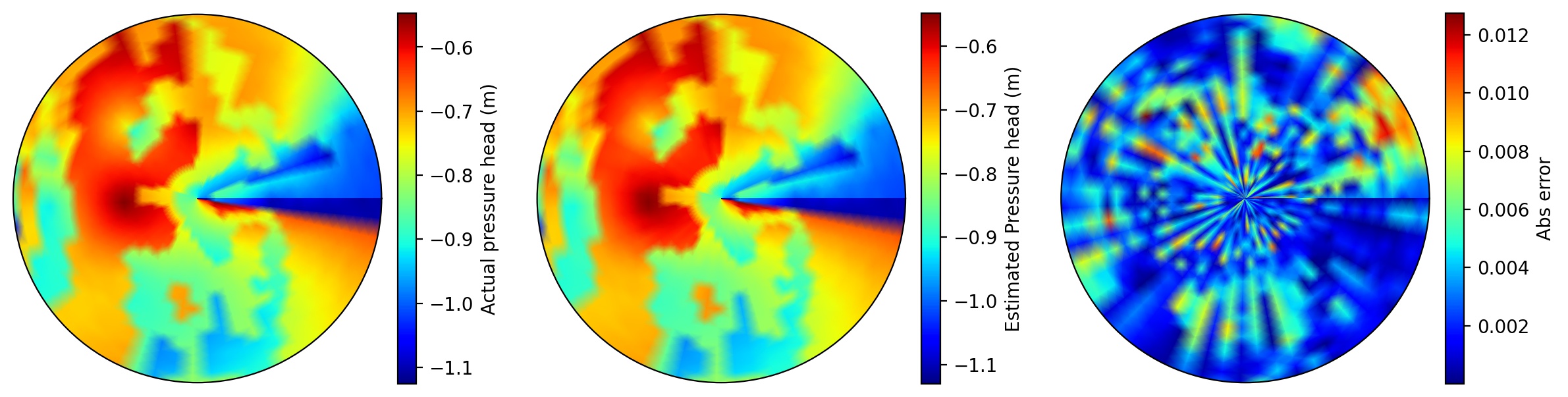}}
\caption{(a) Surface soil moisture map for actual states, estimated states and absolute estimation error at time $t=1$ hr (Left to right) (b) Surface soil moisture map for actual states, estimated states and absolute estimation error at time $t=24$ hr (Left to right) (scenario 1)}
\label{fig:surface_case1}
\end{figure}

\begin{figure}[h]
\centering
\subfigure[]{
\includegraphics[width=1\columnwidth]{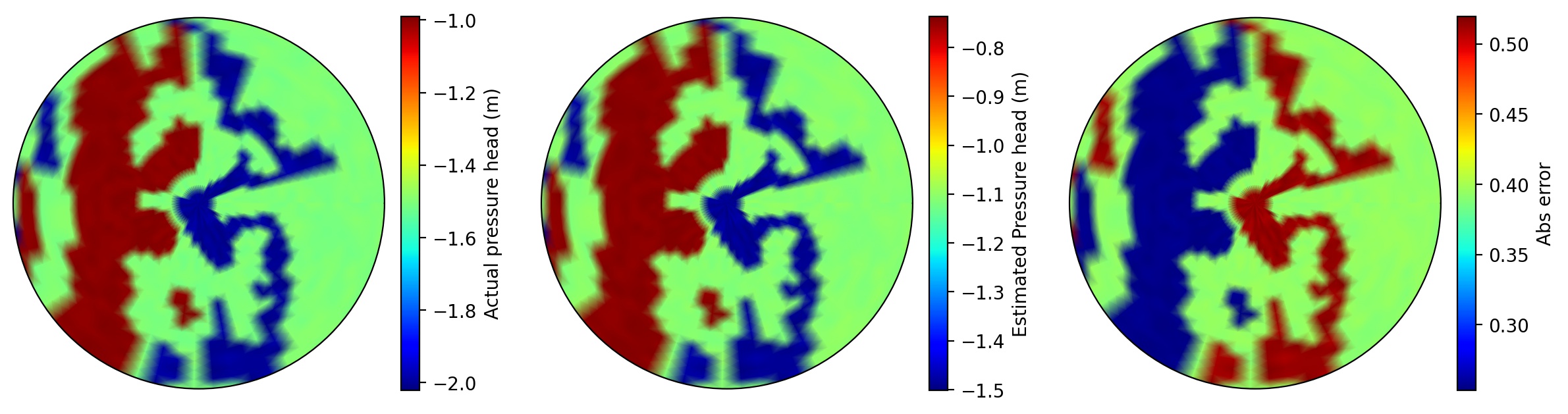}}
\qquad
\subfigure[]{
\includegraphics[width=1\columnwidth]{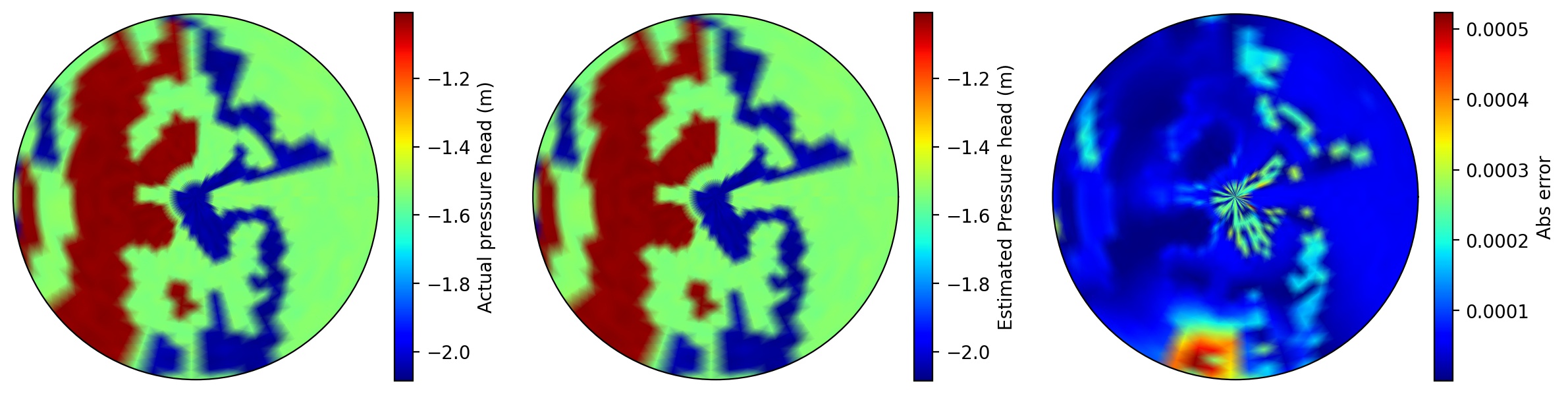}}
\caption{(a) Bottom soil moisture map for actual states, estimated states and absolute estimation error at time $t=1$ hr (Left to right) (b) Bottom soil moisture map for actual states, estimated states and absolute estimation error at time $t=24$ hr (Left to right) (scenario 1)}
\label{fig:bottom_case1}
\end{figure}

In this scenario, we consider the field model without process and measurement noise. The adaptive moving horizon estimation method is applied to estimate the states of the field model. It is assumed that the actual process of the system is known, but the estimator only knows the actual measurement. The initial guess to the adaptive MHE is $\bar{x}_0 = 0.5 \times x_0$, where $x_0$ is the initial condition of the actual system and  $\bar{x}_0$ is the initial condition of the adaptive MHE. The total simulation time is 24 hours. The $Q$ and $R$ values are considered as an identity matrix with diagonal elements as $Q= 2.5e-05\times I_{Nx}$ and $R= 0.01\times I_{Ny}$ respectively. The $P$ matrix is designed based on Equation \ref{eq:P} using the lower and upper bound of the states. The irrigation amount is considered as $0.72$ mm/hr. 
Figure \ref{fig:NC_case1} shows the number of clusters at each time interval and it shows that around 20-130 numbers of states are required each time to represent the system dynamics. Figure \ref{fig:compare_traj_case1} presents the trajectories of some selected states of the actual states and the estimated states. It can be observed that the proposed adaptive moving horizon estimation can track the actual system states accurately. For further analysis, Figure \ref{fig:MSE_case1} shows the MSE between the adaptive MHE and actual system. It can be observed that the MSE converges to a low value after few hours of simulation, which shows the superiority of the proposed adaptive MHE.
Next, the surface soil pressure head for the actual system states, estimated system states, and absolute estimation error at time 1 hr and 24 hr are presented in Figure \ref{fig:surface_case1}. From Figure \ref{fig:surface_case1}(b), it can be observed that the absolute estimation error at time 24 hr is very much lower than the absolute error at time 1 hr (Figure \ref{fig:surface_case1}(a)). In Figure \ref{fig:bottom_case1}, the soil moisture map for the actual system, estimated system states, and absolute estimation error at a depth of 0.5 m is presented. Similar to the surface layer, the estimation error decreases from time $1^{st}$ hr to $24^{th}$ hour. 

\subsection{Scenario 2: noisy case} 
In this scenario, the performance of the adaptive MHE is discussed in the presence of process and measurement noise. The total simulation time is 24 hours, and the initial guess is considered 0.5 times the actual soil moisture. The same P, Q, R values are used as the nominal case. The irrigation amount is $0.72$ mm/hr. The process noise and measurement noise are considered normally distributed noise with zero mean and standard deviation of $6 \times 10^{-3}$ m and $4 \times 10^{-3} $ m. 

\begin{figure}[H]
\centering
\includegraphics[width=1\columnwidth]{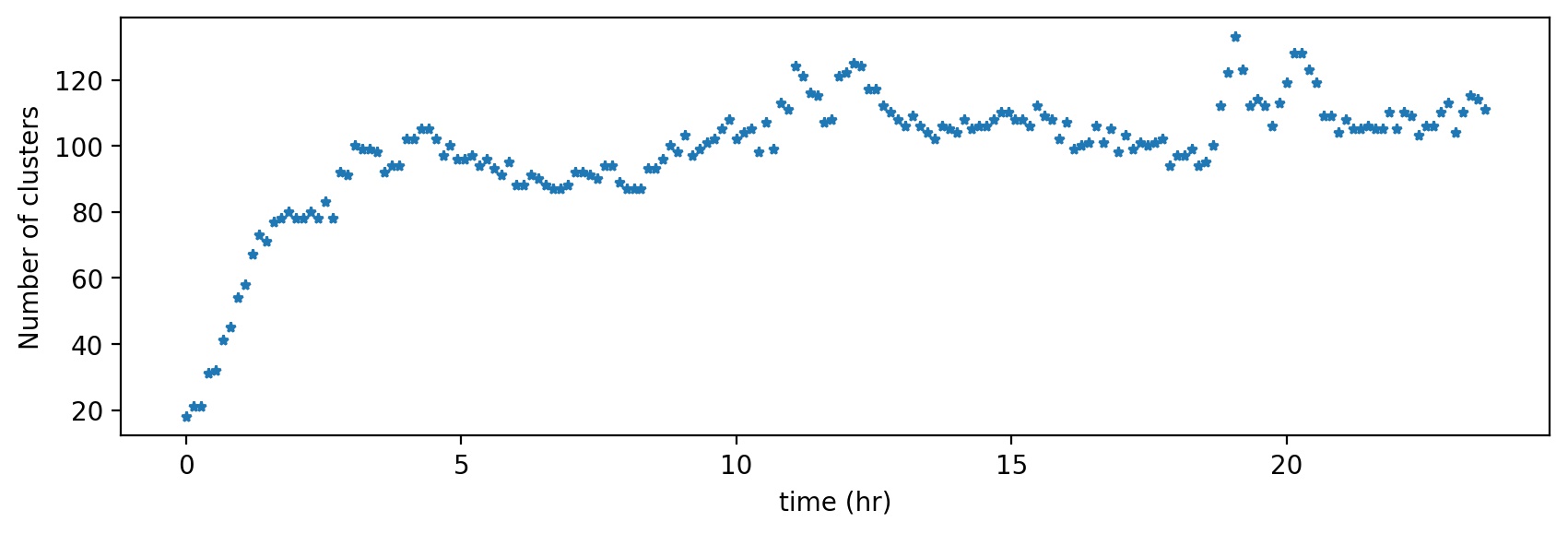}
\caption{Number of clusters for scenario 2}
\label{fig:NC_case2}
\end{figure}

\begin{figure}[H]
\centering
\includegraphics[width=1\columnwidth]{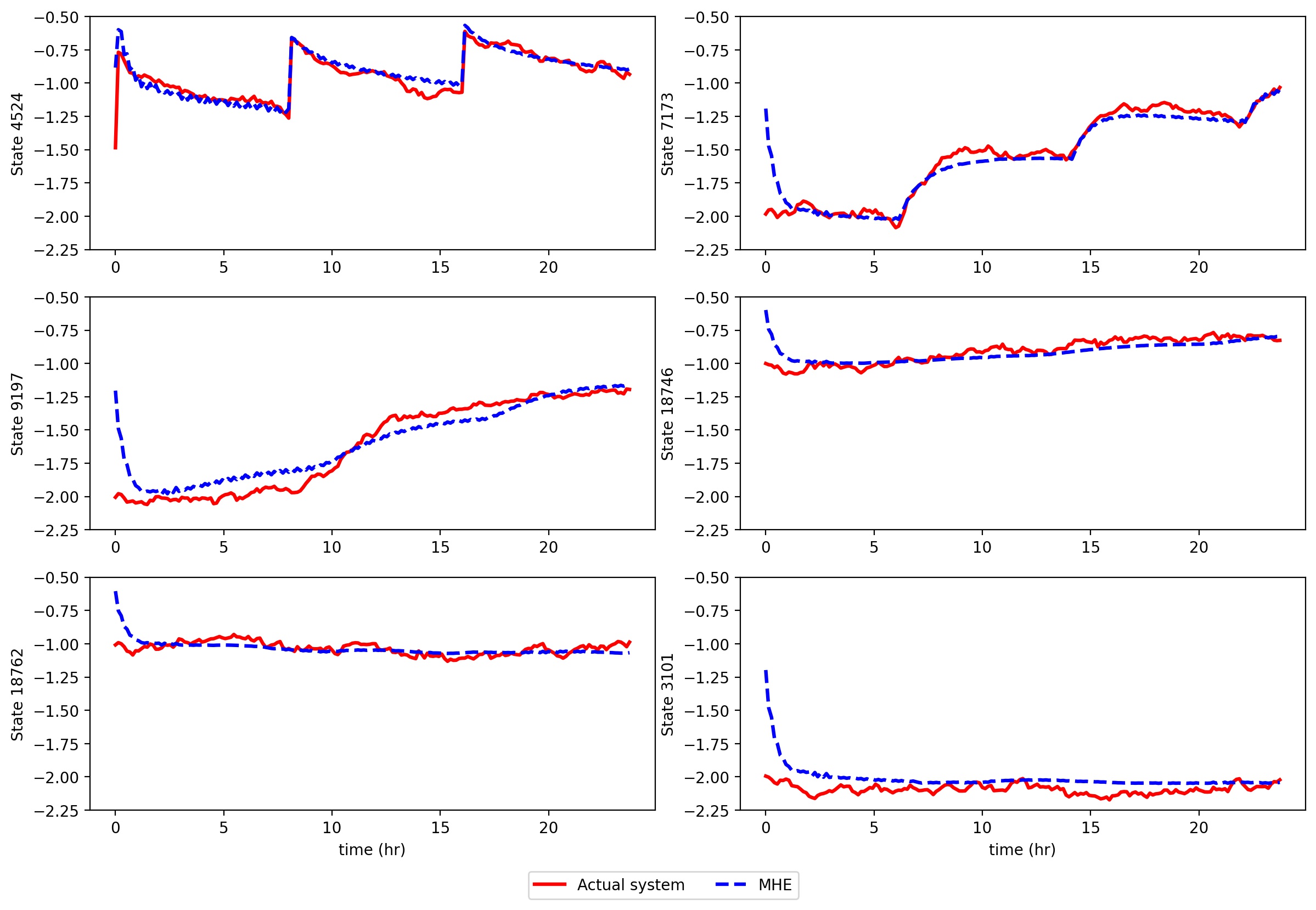}
\caption{Selected state trajectories of the actual states (red solid line), estimated states (blue dotted line) for scenario 2}
\label{fig:compare_traj_case2}
\end{figure}

\begin{figure}[h]
\centering
\includegraphics[width=0.8\columnwidth]{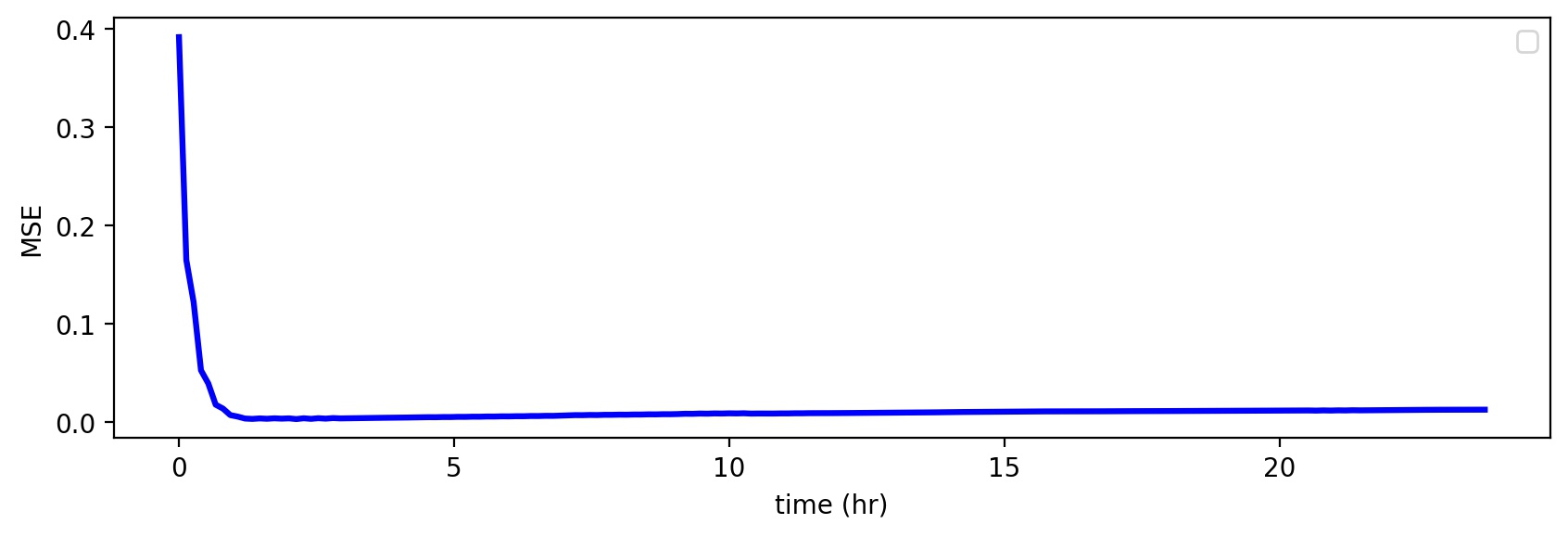}
\caption{Trajectory mean square error for adaptive MHE for scenario 2}
\label{fig:MSE_case2}
\end{figure}

The number of clusters generated for the adaptive MHE is shown in Figure \ref{fig:NC_case2}. Things to note that the tolerance used to generate the clusters is higher than that of the process noise so that the noise does not affect the number of clusters. Figure \ref{fig:compare_traj_case2} shows the trajectories of actual states and estimated states for few selected states. The results show that the estimated states can track the actual trajectory in the presence of noise.
The MSE of estimation error is also presented in Figure \ref{fig:MSE_case2}. The results show that the estimation error converges to a very low value within a few hours of simulation. 

\begin{figure}[h]
\centering
\subfigure[]{
\includegraphics[width=1\columnwidth]{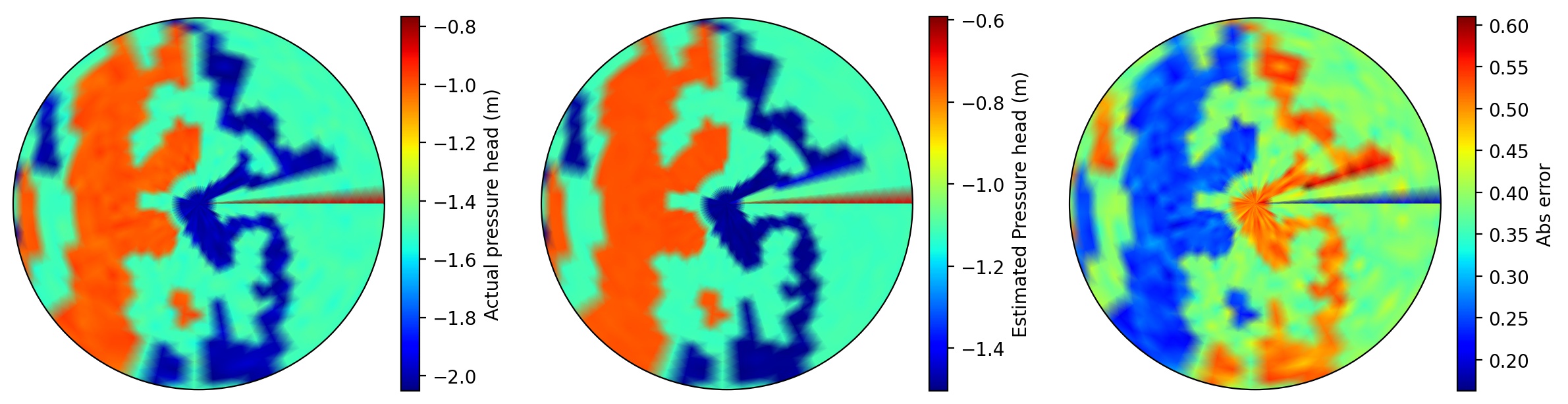}}
\qquad
\subfigure[]{
\includegraphics[width=1\columnwidth]{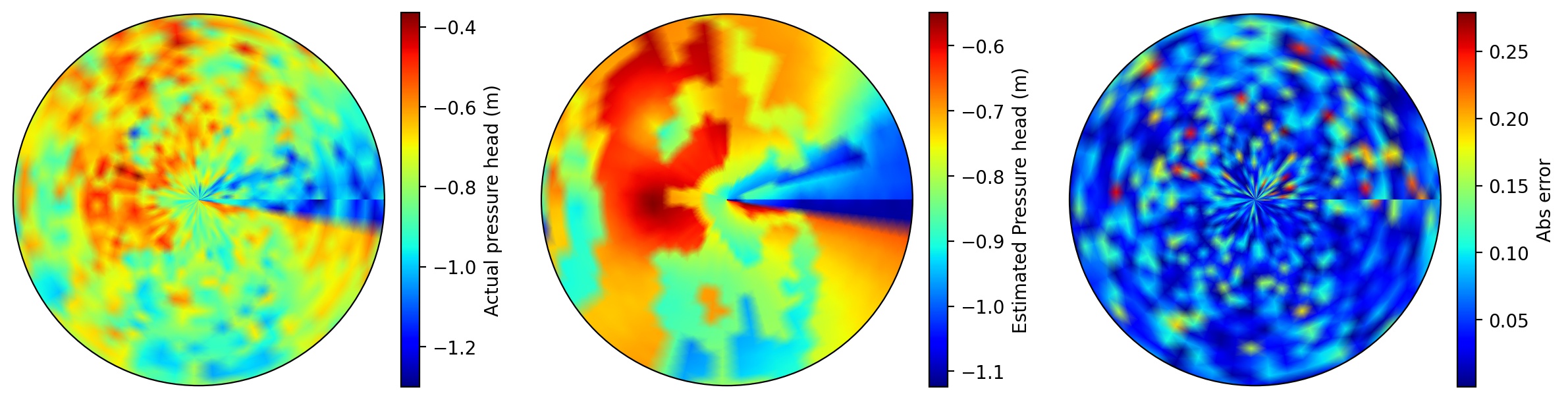}}
\caption{(a) Surface soil moisture map for actual states, estimated states and absolute estimation error at time $t=1$ hr (Left to right) (b) Surface soil moisture map for actual states, estimated states and absolute estimation error at time $t=24$ hr (Left to right) (scenario 2)}
\label{fig:surface_case2}
\end{figure}

\begin{figure}[H]
\centering
\subfigure[]{
\includegraphics[width=1\columnwidth]{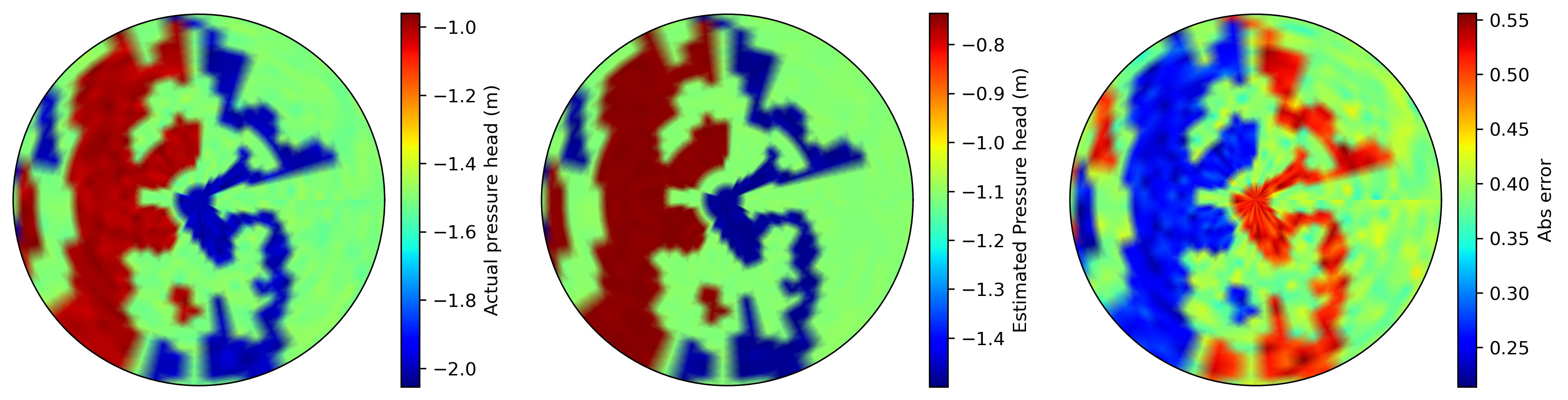}}
\qquad
\subfigure[]{
\includegraphics[width=1\columnwidth]{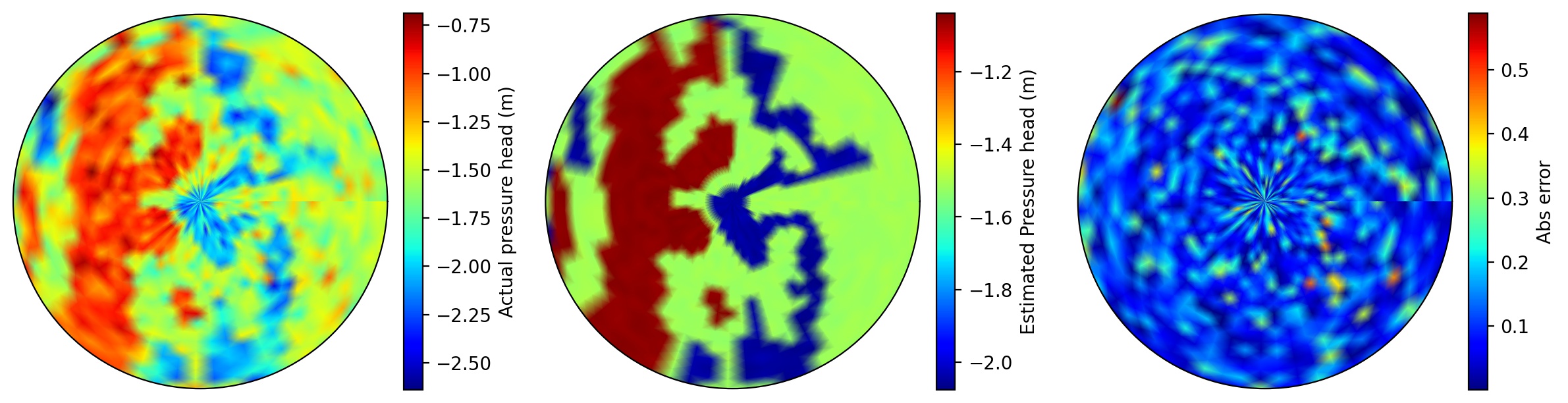}}
\caption{(a) Bottom soil moisture map for actual states, estimated states and absolute estimation error at time $t=1$ hr (Left to right) (b) Bottom soil moisture map for actual states, estimated states and absolute estimation error at time $t=24$ hr (Left to right) (scenario 2)}
\label{fig:bottom_case2}
\end{figure}

For further analysis, the actual system states, estimated system states, and absolute estimation error for the surface nodes at time 1 hr and 24 hr are presented in Figure \ref{fig:surface_case2}. The figure shows that the absolute error at time 24 hr is less than the absolute error at time 1 hr. Similarly, in Figure \ref{fig:bottom_case2}, the soil moisture map for the actual system states, estimated system state, and absolute estimation error at the bottom layer are presented. We can also see a similar trend as the surface layer where the estimation error at $24^{th}$ hr is lesser than the $1^{st}$ hour of simulation. 

\section{Conclusions}
In this article, the problem of higher dimensionality of the agro-hydrological system is addressed. The cylindrical version of the finite difference model is explicitly constructed for a field equipped with a central pivot. The algorithm for structure-preserving adaptive model reduction and adaptive state estimation is proposed. For a small simulated field, the original MHE performance is compared with the proposed MHE method. The results show that the adaptive MHE has a better performance than the original MHE in terms of estimation error and computational cost. The robustness of the adaptive MHE is also presented. The proposed approach has been applied to an real agricultural field under the noisy and nominal case. In both cases, the state estimation shows a satisfactory result and shows the effectiveness of the proposed adaptive model reduction methods. 

\section{Acknowledgment}
Financial support from Alberta Innovates Technology Futures is gratefully acknowledged


\end{document}